\documentclass[twocolumn,showpacs,showkeys,preprintnumbers,amssymb,aps,superscriptaddress,pre]{revtex4}
\usepackage{graphicx}
\usepackage{dcolumn}
\usepackage{bm}
\usepackage{multirow}
\usepackage{longtable}
\usepackage{epsfig}
\usepackage{color}
\usepackage{amsmath}
\usepackage[toc]{appendix}

\begin{document}

\preprint{APS/123-QED}

\title{Magnetization plateau as a result of the uniform and gradual electron doping \\
in a coupled spin-electron double-tetrahedral chain}
\author{Lucia G\'{a}lisov\'{a}}
\email{galisova.lucia@gmail.com}
\affiliation{Department of Applied Mathematics and Informatics, Faculty of Mechanical Engineering, Technical University of Ko\v{s}ice, Letn\'{a} 9, 042 00 Ko\v{s}ice, Slovak Republic}

\date{\today}

\begin{abstract}
The double-tetrahedral chain in a longitudinal magnetic field, whose nodal lattice sites occupied by the localized Ising spins regularly alternate with triangular plaquettes with the dynamics described by the Hubbard model, is rigorously investigated. It is demonstrated that the uniform change of electron concentration controlled by the chemical potential in a combination with the competition between model parameters and the external magnetic field leads to the formation of one chiral and seven non-chiral phases at the absolute zero temperature. Rational plateaux at one-third and one-half of the saturation magnetization can  also be identified in the low-temperature magnetization curves. On the other hand, the gradual electron doping results in eleven different ground-state regions which distinguish from each other by the evolution of the electron distribution during this process. Several doping-dependent magnetization plateaux are observed in the magnetization process as a result of the continuous change of electron content in the model.
\end{abstract}


\maketitle

\section{Introduction}

Magnetization process in one-dimensional (1D) strongly correlated electron systems and quantum spin systems belongs to hot topics of modern condensed matter physics, both from the theoretical and experimental point of view~\cite{Tsv01, Sch04, Rich04}. In particular, the phenomenon of magnetization plateaux in low-temperature magnetization curves is currently a subject of intensive research, because it often reflects an existence of quantum ground states with an intriguing spin arrangement. Many theoretical studies of 1D systems, including various chains~\cite{Hid94, Sak98, Sch02, Oku03, Str03, Hon04, Can06, Can09, Ant09, Roj11, Lis13, Ana13, Ana14, Str14, Abg15, Ver16}, ladders~\cite{Cab97, Cab98, Oka02, Jap06, Iva06, Mic10, Kar15} and/or other related models~\cite{Fou06, Oku12}, indicate a presence of the magnetization plateaux at certain {\it rational} fractions of the saturation magnetization. The occurrence of these plateaux is subject to a validity of the quantized condition known as Oshikawa-Yamanaka-Affleck (OYA) rule~\cite{Osh97},
\begin{eqnarray}
\label{eq:OYA}
p(m_{sat} - m) = {\rm integer},
\end{eqnarray}
where $p$ is the period of the ground state, while $m_{sat}$ and $m$ are, respectively, the saturation and total magnetization per elementary unit of the ground state. It should be noted that the criterion~(\ref{eq:OYA}) is a necessary but not a sufficient condition for the {\it rational} magnetization plateau formation. Experimentally, {\it  rational} values of the saturation magnetization can be observed in numerous insulating materials, such as the azurite Cu$_3$(CO$_3$)$_2$(OH)$_2$~\cite{Kik04, Oht04, Kik05}, Cu(3-Clpy)$_2$(N$_3$)$_2$~\cite{Hag01, Hag03} Cu$_3$(P$_2$O$_6$OM)$_2$~(M=H,\,\,P)~\cite{Has06, Has07}, NH$_4$CuCl$_3$~\cite{Shi98}, Cu$_3$Cl$_6$(H$_2$O)$_2\cdot2$H$_8$C$_4$SO$_2$~\cite{Ish00}, etc.

In context of the magnetization process of 1D systems, magnetization curves with plateaux at non-trivial {\it irrational} values can also be found both experimentally~\cite{Vis09, Heu10} and theoretically~\cite{Bel14, Oha15, Cab00, Puj02, Lam06, Cab01, Cab02, Rou06, Rou07, Str16}. It was confirmed that this striking phenomenon is not in conflict with the OYA criterion~(\ref{eq:OYA}) and may be a direct result of different Land\'e $g$-factors of magnetic particles~\cite{Bel14, Oha15, Vis09, Heu10,Tor17}, quenched disorder~\cite{Cab00, Puj02, Lam06} or doping by itinerant particles~\cite{Cab01, Cab02, Rou06, Rou07, Str16}. In the last case, a change of the particle concentration may allow one to tune the value and position of the magnetization plateau in a controlled way. Of particular interest is the mechanism leading to a continuous variation of the magnetization plateau with the particle filling. A valuable feature of this mechanism is that the controlled doping allows one to move the doping-dependent plateaux to lower magnetic fields~\cite{Cab01, Cab02, Rou06, Rou07}. This makes the afore-mentioned mechanism very attractive for experiments.

In the present work, we will examine ground-state properties and the low-temperature magnetization process of a coupled spin-electron double-tetrahedral chain, in which nodal lattice sites occupied by the localized Ising spins of the magnitude $\sigma=1/2$ regularly alternate with triangular plaquettes available for mobile electrons. The magnetic structure of this 1D lattice is experimentally realized in the copper-based polymeric chain Cu$_3$Mo$_2$O$_9$~\cite{Has08, Kur11, Mat12, Kur14}. As has been shown in our recent works~\cite{Gal15a, Gal15b, Gal15c, Gal17}, the mixed spin-electron system with the lattice topology of the double-tetrahedral chain represents an excellent prototype model, which allows the exact investigation of many interesting physical phenomena such as quantum non-chiral and/or uncommon chiral ground states, the rational magnetization plateau, the abnormally narrow and high low-temperature peak of the specific heat and very abrupt (almost discontinuous) thermal variations of the entropy and sublattice magnetization caused by a difference in ground-state degeneracies, as well as the enhanced magnetocaloric effect. The main purpose of this paper is to explore in detail an effect of the uniform and gradual electron doping on the ground-state properties of the model and the plateau creation in the magnetization process in order to provide a deeper insight into doping-dependent magnetization plateaux.

The paper is organized as follows. In Sec.~\ref{sec:2} we present the magnetic structure of the model and briefly mention the computational idea which leads to the rigorous solution of this model. Secs.~\ref{sec:3} and~\ref{sec:4} present numerical results for the ground state and low-temperature magnetization process during the uniform and gradual electron doping, respectively. The last section (Sec.~\ref{sec:5}) contains a~summary of the most interesting results and the conclusion. The paper ends with three appendices which include the complete set of electron basis states of the orthogonal Hilbert subspaces corresponding to all possible numbers of mobile electrons in the triangular plaquette and energy eigenvalues of one double-tetrahedron (Appendix~\ref{app:1}), the state vectors corresponding to the $k$th triangular plaquette which are parts of the eigenvectors of ground-state phases appearing during the uniform electron doping (Appendix~\ref{app:2}) and the list of analytical expressions for first-order phase transitions between these ground states (Appendix~\ref{app:3}).

\section{Spin-electron double-tetrahedral chain}
\label{sec:2}
\begin{figure}[t!]
\begin{center}
\hspace{0.25cm}
\includegraphics[width=1.0\columnwidth]{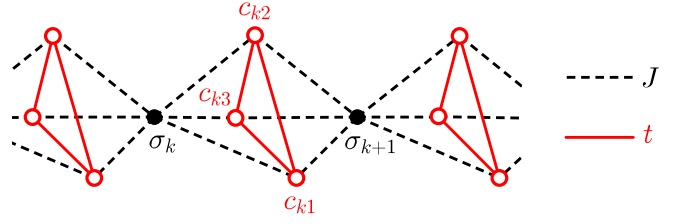}
\vspace{-0.45cm}
\caption{\small (Color online) A part of the spin-electron double-tetrahedral chain. Full circles denote nodal sites occupied by the localized Ising spins $\sigma = 1/2$ and empty circles forming triangular clusters are available to mobile electrons.}
\label{fig1}
\end{center}
\vspace{0.0cm}
\end{figure}

We consider a magnetic system with a lattice topology of the double-tetrahedral chain, in which nodal lattice sites occupied by the localized Ising spins regularly alternate with triangular plaquettes consisting of three equivalent lattice sites available for mobile electrons (see Fig.~\ref{fig1}). Assuming $N$ nodal sites, the total Hamiltonian of this spin-electron model can be expressed as a sum of $N$ cluster Hamiltonians $\hat{\cal H}_k$:
\begin{eqnarray}
\label{eq:H}
\hat{{\cal H}} \!\!&=&\!\! \sum_{k = 1}^N\hat{{\cal H}}_k,
\\
\label{eq:Hk}
\hat{{\cal H}}_k \!\!&=&\!\! -t\!\!\!\!\sum_{\gamma \in\{\uparrow, \downarrow\}}\!\sum_{j=1}^3 \!\big(\hat{c}_{kj,\gamma}^{\dag}\hat{c}_{k(j+1){_{\rm mod 3}},\gamma} \!+ {\rm H.c.}\big)
\nonumber\\
\!\!&+&\!\!\frac{J}{2} (\sigma_{k}^{z} + \sigma_{k+1}^{z}) \sum_{j = 1}^{3}\,(\hat{n}_{kj,\uparrow} - \hat{n}_{kj,\downarrow}) + U \sum_{j = 1}^{3}\hat{n}_{kj,\uparrow}\hat{n}_{kj,\downarrow}
\nonumber\\
\!\!&-&\!\! \mu \sum_{j = 1}^{3}\,(\hat{n}_{kj,\uparrow} + \hat{n}_{kj,\downarrow}) - \frac{H_{e}}{2} \sum_{j = 1}^{3}\,(\hat{n}_{kj,\uparrow} - \hat{n}_{kj,\downarrow})
\nonumber\\
\!\!&-&\!\!
\frac{H_{I}}{2} (\sigma_{k}^{z} + \sigma_{k+1}^{z}).
\end{eqnarray}
Each cluster Hamintonian~(\ref{eq:Hk}) contains all the interaction terms connected to one triangular plaquette and its two surrounding nodal lattice sites. Parameters $\hat{c}_{kj,\gamma}^{\dag}$, $\hat{c}_{kj,\gamma}$ represent usual fermionic creation and annihilation operators for mobile electrons with the spin $\gamma \in\{ \uparrow, \downarrow\}$ that occupy the $j$th site of the triangular plaquette in $k$th position, $\hat{n}_{kj,\gamma} = \hat{c}_{kj,\gamma}^{\dag}\hat{c}_{kj,\gamma}$ denotes the respective number operator and $\sigma_{k}^{z}= \pm 1/2$ labels the Ising spin placed at the $k$th nodal site. The hopping parameter $t>0$ takes into account the kinetic energy of mobile electrons in triangular plaquettes, $J$ stands for the Ising-type coupling between electrons and their nearest Ising neighbors, $U>0$ represents the on-site Coulomb repulsion between two electrons at the same plaquette site and $\mu$ is the chemical potential, which allows one to tune the electron content in the studied system. The last two terms in Eq.~(\ref{eq:Hk}) represent the Zeeman's energies of mobile electrons and the Ising spins in a presence of the magnetic fields $H_{e}$ and $H_{I}$, respectively. Finally, we have introduced the modulo 3 operation into the first term of  cluster Hamiltonian~(\ref{eq:Hk}) in order to ensure the periodic boundary condition $\hat{c}_{k4,\gamma}^{\dag} = \hat{c}_{k1,\gamma}^{\dag}$ ($\hat{c}_{k4,\gamma} = \hat{c}_{k1,\gamma}$) for the three-site electron subsystem.

The cluster Hamiltonian~(\ref{eq:Hk}) can be alternatively written in terms of the
operators $\hat{n}_{k} = \sum_{j = 1}^{3}(\hat{n}_{kj,\uparrow}+\hat{n}_{kj,\downarrow})$ and $\hat{S}_k^z = \sum_{j = 1}^{3}(\hat{n}_{kj,\uparrow}-\hat{n}_{kj,\downarrow})/2$. The former operator determines the total number $n_k$ of mobile electrons in the triangular plaquette of $k$th double-tetrahedron, while the latter one specifies the total spin $S_k^z$ of this plaquette in the $z$th direction. It is worth to note that the Hamiltonian~(\ref{eq:Hk}) can be written as $64\times64$ matrix, if one considers a variable number of electrons from zero up to six in the triangular plaquette. This matrix has the block-diagonal form with regard to a validity of the commutation relations $[\hat{{\cal H}}_k, \hat{n}_{k}] = 0$, $[\hat{{\cal H}}_k, \hat{S}_k^z] = 0$. Individual matrix blocks of~(\ref{eq:Hk}) correspond to the orthogonal Hilbert subspaces with different but fixed number of electrons $n_k$ and distinct values of the total spin $S_k^z$. The eigenvalues of these blocks can be analytically found by applying the so-called basis of wavelets~\cite{Lul03, Jak14} on the orbits ${\cal O}_{|f_{kj}^{\rm ini}\rangle}$ of individual Hilbert subspaces as the aftermath of the cyclic translational symmetry of triangular plaquettes (see Appendix A). It should be mentioned that each orbit ${\cal O}_{|f_{kj}^{\rm ini}\rangle}$ is well defined by the initial electron configuration $|f_{kj}^{\rm ini}\rangle = \underbrace{\hat{c}_{kj,\gamma_j}^{\dag}\hat{c}_{kp,\gamma_p}^{\dag}\ldots\hat{c}_{kq,\gamma_q}^{\dag}}_{n_k}|0\rangle$ ($j, p,\ldots, q\in \{1, 2, 3\}$, $\gamma_j, \gamma_p,\ldots, \gamma_q\in\{\uparrow,\downarrow\}$, $|0\rangle$ labels the vacuum state) and consists of electron configuration(s) satisfying
\begin{eqnarray}
{\rm c}_3 |f_{kj}^{\rm ini}\rangle \!\!&=&\!\! \hat{c}_{k(j+1)_{{\rm mod} 3},\gamma_j}^{\dag}\hat{c}_{k(p+1)_{{\rm mod} 3},\gamma_p}^{\dag}\ldots\hat{c}_{k(q+1)_{{\rm mod} 3},\gamma_q}^{\dag}|0\rangle,
\nonumber\\
{\rm c}_3^2 |f_{kj}^{\rm ini}\rangle \!\!&=&\!\! {\rm c}_3\left({\rm c}_3 |f_{kj}^{\rm ini}\rangle\right)
\nonumber \\
\!\!&=&\!\!
\hat{c}_{k(j+2)_{{\rm mod} 3},\gamma_j}^{\dag}\hat{c}_{k(p+2)_{{\rm mod} 3},\gamma_p}^{\dag}\ldots\hat{c}_{k(q+2)_{{\rm mod} 3},\gamma_q}^{\dag}|0\rangle.
\nonumber
\end{eqnarray}
In above, ${\rm c}_3$ is an element of the cyclic translation group~${\rm C}_3$.

A complete set of eigenvalues of the cluster Hamiltonian~(\ref{eq:Hk}) can be employed for a calculation of the grand free energy~$\Omega$ per elementary unit:
\begin{eqnarray}
\label{eq:Omega}
\Omega\!\!&=&\!\!
-k_{\rm B} T\lim_{N\to\infty}\frac{1}{N}\ln\Big(\sum_{\{\sigma_k\!\}}\mathrm{Tr}\,e^{-\beta\hat{{\cal H}}}\Big)
\nonumber\\
\!\!&=&\!\! -k_{\rm B} T\lim_{N\to\infty}\frac{1}{N}\ln\Big(\sum_{\{\sigma_k\!\}}\prod_{k=1}^N\mathrm{Tr}_k\,e^{-\beta\hat{{\cal H}}_k}\Big)
\nonumber\\
\!\!&=&\!\!
-k_{\rm B} T\lim_{N\to\infty}\frac{1}{N}\ln\Big(\sum_{\{\sigma_k\!\}}\prod_{k=1}^N\sum_{l=1}^{64}e^{-\beta E_{kl}} \Big).
\end{eqnarray}
In above, $\beta=1/(k_{\rm B}T)$ represents the inverse temperature ($k_{\rm B}$ is Boltzmann's constant and $T$ is the absolute temperature), $\sum_{\{\sigma_k \}}$ denotes a summation over all possible states of the localized Ising spins and $\mathrm{Tr}$ labels for a trace over the degrees of freedom of all mobile electrons in the system. The trace $\mathrm{Tr}$ can be factorized into a product of $N$ traces $\mathrm{Tr}_k$ running over degrees of freedom of electrons from the $k$th plaquette due to a validity of the commutation relation between different cluster Hamiltonians $[\hat{{\cal H}}_k, \hat{{\cal H}}_{k'}] = 0$. It is worthy to note that the grand free energy~$\Omega$ is valuable for investigating all physical quantities which may be helpful in understanding the magnetic behavior of the considered model. A rigorous expression for~$\Omega$ may be obtained by two independent analytical methods. The first method is the standard transfer-matrix technique~\cite{Kra44, Bax82, Str15}, while the second one lies in a combination of the generalized decoration-iteration mapping transformation~\cite{Fis59, Syo72, Str10} with the known analytical formula for the partition function of the spin-1/2 Ising chain in the longitudinal field~\cite{Kra44, Bax82, Str15}. The reader can find computational details of both methods in our recent works~\cite{Gal15a, Gal15b, Gal17}, which deal with some particular variants of the model with the conserved electron concentration.

\section{Uniform electron doping}
\label{sec:3}

In this section, we will discuss the ground-state arrangement and low-temperature magnetization process during an electron doping which results in the same electron content in triangular plaquettes. Due to some fundamental differences between magnetic properties of models with distinct nature (sign) of the spin-electron couplings~\cite{Gal15a, Gal15b, Gal17, Gal15c}, our attention will be restricted on the particular case of the ferromagnetic interaction $J<0$ hereafter. The same magnetic fields acting on the Ising spins and mobile electrons $H=H_I=H_e$ will also be assumed in order to reduce the total number of free parameters.

\subsection{Ground state}
\label{subsec:3A}

To find all possible spin-electron arrangements at the zero temperature a comprehensive analysis of all eigenvalues $E_{kl}$ listed in Appendix~A has to be performed by considering four available spin combinations of the Ising pair $\sigma_k$, $\sigma_{k+1}$. The gained lowest-energy eigenstates can be then extended to the whole model due to the commuting character of the cluster Hamiltonian~(\ref{eq:Hk}). In this manner, eight different phases can be observed at $T=0$ as a result of the mutual competition between model parameters and the external magnetic field. The spin-electron arrangements peculiar to these phases are unambiguously determined by the following eigenvectors and energies:
\begin{eqnarray}
\label{eq:S_0,2,4,6}
|{\rm S}_{a}\rangle \!\!&=&\!\!  \left\{      \begin{tabular}{l}
                                                            ${\displaystyle{\prod_{k=1}^N}}$  $|(\downarrow)\!\uparrow\rangle_{\sigma_k}$
                                                            $\!\otimes\!$
                                                            $|n_{k}=a,S_{k}^{z}=0\rangle$,\, $H = 0$
                                                            \\
                                                            ${\displaystyle{\prod_{k=1}^N}}$  $|\!\uparrow\rangle_{\sigma_k}$ $\!\otimes\!$
                                                            $|n_{k}=a,S_{k}^{z}=0\rangle$\hspace{3mm}  ,\, $H > 0$
                                                     \end{tabular}
                                        \right.\!\!,
                                        \nonumber\\
E_{a} \!\!&=&\!\! -\,\frac{NH}{2} -\frac{Na}{16}\bigg[16\mu -(a+2)U -2(a-6)t
\nonumber\\
\!\!&&\!\! - \,(a-6)\sqrt{(U+2t)^2+32t^2}\,\bigg], \, a=\!\{0,2,4,6\};
\\[2mm]
\label{eq:S_1,5}
|{\rm S}_{b}\rangle \!\!&=&\!\! \left\{      \begin{tabular}{l}
                                                            ${\displaystyle{\prod_{k=1}^N}}$
                                                            $|\!\downarrow\rangle_{\sigma_k}$ $\!\otimes\!$
                                                            $\Big|n_{k}=b,S_{k}^{z}=-\dfrac{1}{2}\Big\rangle$,\,
                                                            $H = 0$
                                                            \\
                                                            ${\displaystyle{\prod_{k=1}^N}}$  $|\!\uparrow\rangle_{\sigma_k}$ $\!\otimes\!$
                                                            $\Big|n_{k}=b,S_{k}^{z}=\dfrac{1}{2}\Big\rangle$\hspace{4mm}  \hspace{-2.2mm},\,\, $H \geq 0$
                                                     \end{tabular}
                                        \right.\!\!,
                                        \nonumber\\
E_{b} \!\!&=&\!\! \frac{NJ}{2} - NH -\frac{N}{2}\big[2b\mu
\nonumber\\
& & \hspace{1.45cm}{} - (b-1)U + 4t \big],\quad b=\!\{1,5\};
\\[2mm]
\label{eq:S_3}
|{\rm S}_{3}\rangle \!\!&=&\!\! \left\{ \begin{tabular}{l}
                                                            ${\displaystyle{\prod_{k=1}^N}}$
                                                            $|\!\downarrow\rangle_{\sigma_k}$ $\!\otimes\!$
                                                            $\Big|n_{k}=3,S_{k}^{z}=-\dfrac{3}{2}\Big\rangle$,\,
                                                            $H = 0$
                                                            \\
                                                            ${\displaystyle{\prod_{k=1}^N}}$  $|\!\uparrow\rangle_{\sigma_k}$ $\!\otimes\!$
                                                            $\Big|n_{k}=3,S_{k}^{z}=\dfrac{3}{2}\Big\rangle$\hspace{7mm}  \hspace{-6mm},\, $H \geq 0$
                                                     \end{tabular}
                                               \right.\!\!,
                                                \nonumber\\
E_{3} \!\!&=&\!\! \frac{N}{2}\left(3J - 4H - 6\mu\right);
\\[2mm]
\label{eq:S_3chiral}
|{\rm \widetilde{S}}_{3}\rangle \!\!&=&\!\! \left\{ \begin{tabular}{l}
                                                            ${\displaystyle{\prod_{k=1}^N}}$
                                                            $|\!\downarrow\rangle_{\sigma_k}$ $\!\otimes\!$
                                                            $\Big|n_{k}=3,S_{k}^{z}=-\dfrac{1}{2}\Big\rangle_{L, R}$
                                                            ,\,
                                                            $H = 0$
                                                            \\
                                                            ${\displaystyle{\prod_{k=1}^N}}$  $|\!\uparrow\rangle_{\sigma_k}$ $\!\otimes\!$
                                                            $\Big|n_{k}=3,S_{k}^{z}=\dfrac{1}{2}\Big\rangle_{ L, R}$\hspace{4mm},\, $H \geq 0$
                                                     \end{tabular}
                                               \right.\!\!,
                                                \nonumber
\end{eqnarray}
\begin{eqnarray}
\widetilde{E}_{3} \!\!&=&\!\! \frac{NJ}{2} - NH -\frac{N}{3}\big[9\mu -2U {}
\nonumber\\
& & \hspace{1.6cm}{}+ 6\sqrt{U^{2}+27t^{2}}\cos\left(\phi_{U,\,t}\right)\big]. {}
\end{eqnarray}
In above, the subscripts $a$, $b$ and $3$ specify a total number (concentration) of electrons in the $k$th triangular plaquette and $\phi_{U,\,t}= \frac{1}{3}\arctan\!\left(\frac{9t}{U^{3}}\sqrt{U^{4}+27U^{2}t^{2}+243t^{4}}\right)$. The product~$\prod_{k=1}^{N}$ runs over elementary clusters, the state vector $|\!\uparrow\rangle_{\sigma_k}$ ($|\!\downarrow\rangle_{\sigma_k}$) determines the spin state $\sigma_k^z=1/2$ ($-1/2$) of the Ising spin located at the $k$th nodal site, while $|n_{k}, S^{z}_{k}\rangle$ and $|n_{k}, S^{z}_{k}\rangle_{L\!,R}$ refer, respectively, the non-chiral and chiral eigenstates of $n_{k}$ electrons with the total spin~$S^{z}_{k}$ in the $k$th plaquette. Analytical expressions of the state vectors $|n_{k}, S^{z}_{k}\rangle$, $|n_{k}, S^{z}_{k}\rangle_{L\!,R}$ are too cumbersome for some electron concentrations, therefore the reader finds them in Appendix~B.

It can be easily deduced from Eqs.~(\ref{eq:S_0,2,4,6})--(\ref{eq:S_3chiral}) that an arrangement of the Ising spins depends on a presence/absence of the magnetic field and the total spin of triangular plaquettes. To be specific, the phases S$_{a}$, where triangular plaquettes have the zero total spin, because they are either empty, fully-filled or partly occupied by two or four electrons in a quantum superposition of several intrinsic antiferromagnetic and non-magnetic states, exhibit a kinetically-driven frustration of the Ising spins at $H=0$. An arbitrary non-zero field cancels this macroscopic degeneracy, because it forces the Ising spins to align into its direction. On the other hand, all the Ising spins may occupy either the spin state $\sigma_k^z=-1/2$ or $1/2$ if the phases S$_{b}$, S$_{3}$ and $\widetilde{\rm S}_{3}$ occur in the zero-field ground state. Mobile electrons in triangular plaquettes also choose between two possible quantum states with the total spins $S_{k}^{z}= -1/2$ and $1/2$ (in S$_{b}$ and $\widetilde{\rm S}_{3}$) or between two classical ferromagnetic states corresponding to the total spins $S_{k}^{z}= -3/2$ and $3/2$ (in S$_{3}$) in order to preserve the spontaneous ferromagnetic ordering with the Ising neighbors. Any non-zero magnetic field lifts this two-fold degeneracy of the phases S$_{b}$, S$_{3}$, $\widetilde{\rm S}_{3}$. Moreover, one can also find a peculiar field-independent macroscopic degeneracy if the phase $\widetilde{\rm S}_{3}$ constitutes the ground state. This degeneracy relates to two possible chiral degrees of freedom ($L$eft- and $R$ight-hand) of mobile electrons in each triangular plaquette. It should be said that four of the eight observed phases have already been identified in some particular variants of the model with the conserved electron content in triangular plaquettes. Namely, the phase S$_1$ represents the ground state of the model with the one-sixth filling of plaquettes~\cite{Gal15a}, S$_2$ is one of two possible phases constituting the zero-temperature phase diagram of the model with the one-third electron filling~\cite{Gal15c} and the phases S$_3$, $\widetilde{\rm S}_{3}$ have already been detected in the ground state of the model with half-filled plaquettes~\cite{Gal17}.

\begin{figure*}[t!]
\vspace{0.25cm}
    \includegraphics[width = 0.45\textwidth]{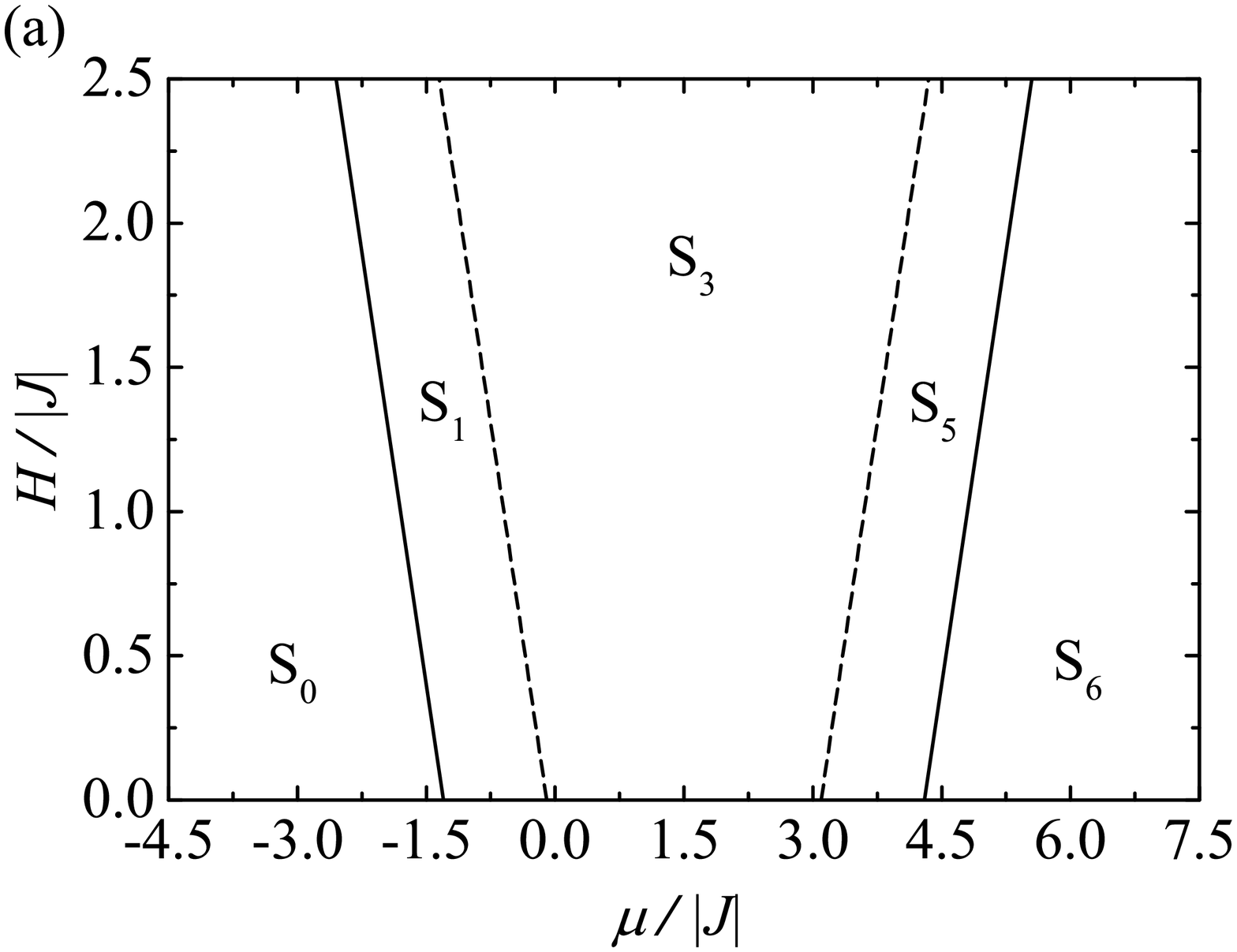}
     \hspace{0.0cm}
    \includegraphics[width = 0.45\textwidth]{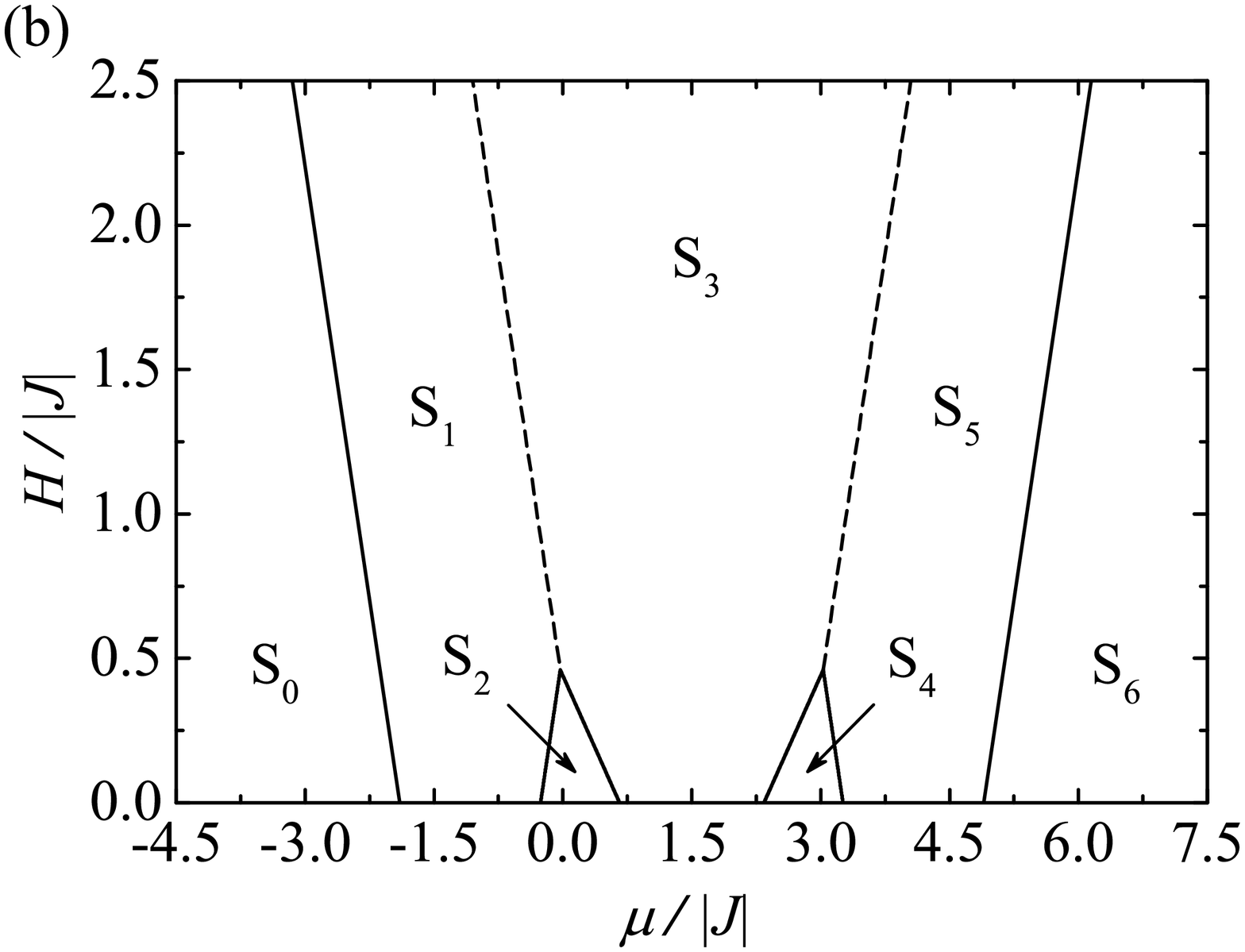}
    \\[0.25cm]
    \includegraphics[width = 0.45\textwidth]{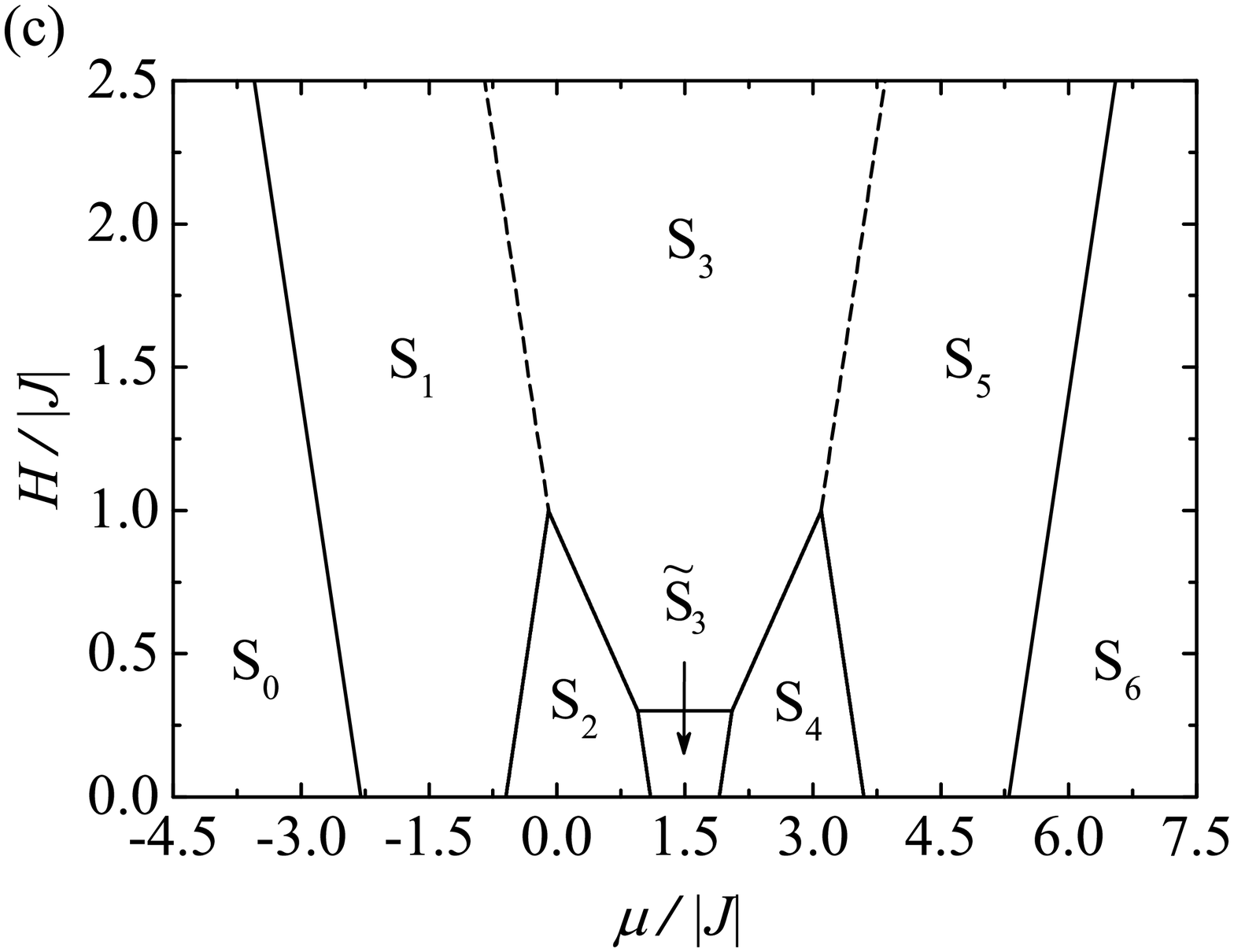}
     \hspace{0.0cm}
    \includegraphics[width = 0.45\textwidth]{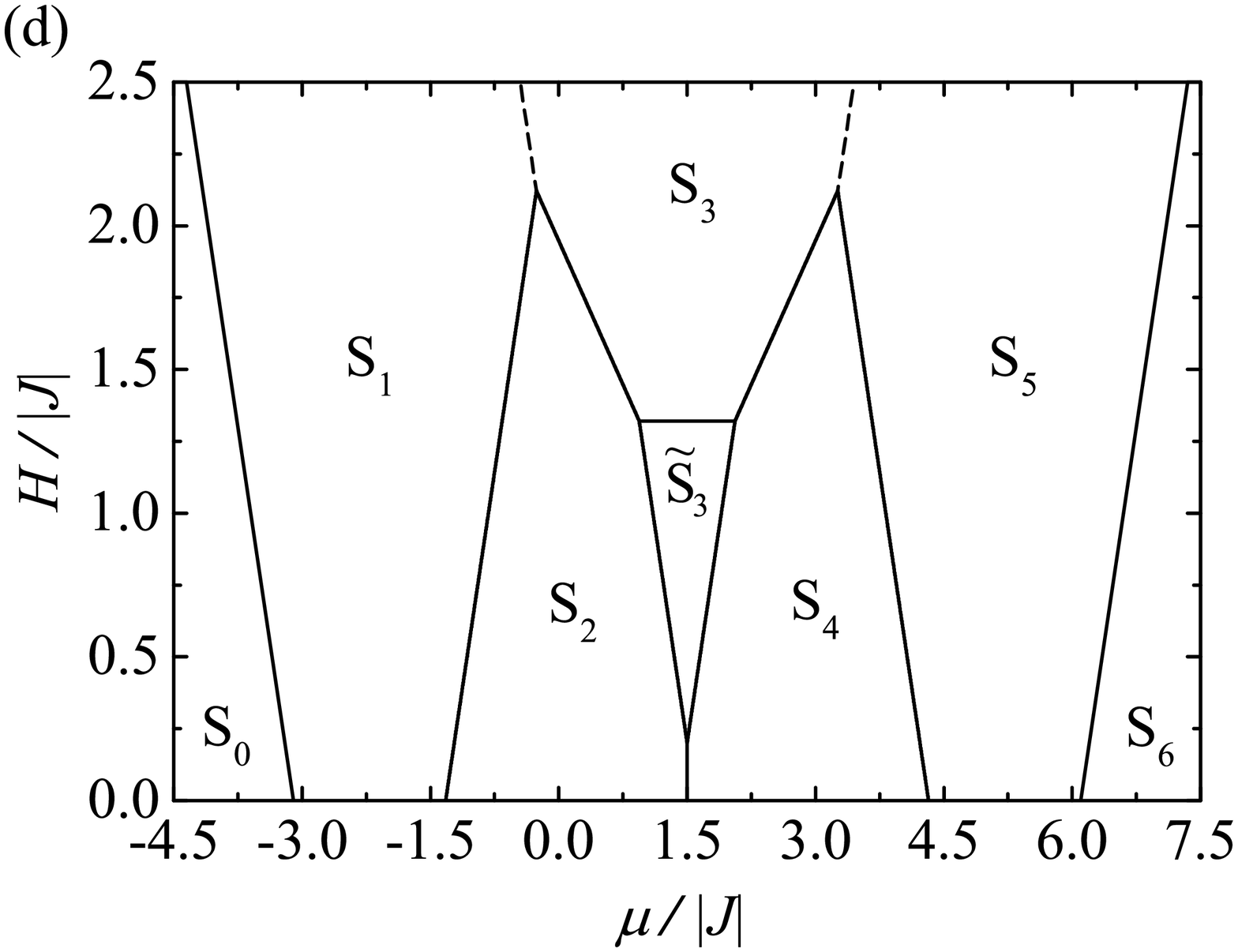}
  \caption{\small Ground-state phase diagrams of the model~(\ref{eq:H}) in the $\mu-H$ plane for the fixed Coulomb term $U/|J| = 3$ and four representative values of the kinetic parameter (a)~$t/|J| = 0.4$, (b)~$t/|J| = 0.7$, (c)~$t/|J| = 0.9$, (d)~$t/|J| = 1.3$.}
  \label{fig2}
\end{figure*}

To illustrate stability regions of the ground states~(\ref{eq:S_0,2,4,6})--(\ref{eq:S_3chiral}), several zero-temperature diagrams are depicted in Fig.~\ref{fig2} for the fixed Coulomb term $U/|J|= 3$ and a few representative values of the hopping parameter $t/|J|$. Obviously, four different ground-state topologies are possible in the $\mu-H$ plane. If the hopping parameter is smaller than the boundary value
\begin{eqnarray}
t_{b1} = -\frac{U}{18} + \frac{1}{18}\sqrt{(U - 6J)^2 - 24UJ},
\label{eq:tb1}
\end{eqnarray}
the $\mu-H$ plane is divided into five regions corresponding to five different phases [see Fig.~\ref{fig2}(a)]: S$_{0}$ and S$_{6}$, where the system is broken up into a set of $N$ non-interacting Ising spins due to zero effective interactions mediated by either empty (in S$_{0}$) or fully-filled non-magnetic (in S$_{6}$) plaquettes; S$_{3}$, in which each plaquette site is occupied by a single electron due to its polarization into the magnetic-field direction; and S$_{1}$, S$_{5}$, which are characterized by a single and five electrons per plaquette, respectively. We note that the phases S$_{0}$, S$_{1}$ represent mirror images of the phases S$_{6}$, S$_{5}$ on account of the fact that empty plaquette sites can be replaced by doubly occupied ones. If the reverse condition $t>t_{b1}$ is satisfied, other two novel phases S$_{2}$, S$_{4}$ can be observed in the ground state  [see Figs.~\ref{fig2}(b)--(d)]. Both these phases are characterized by the zero total spin of triangular plaquettes, since mobile electrons underlie a quantum superposition of six antifferomagnetic and three non-magnetic states [see the state vectors~(\ref{eq:n2}) and~(\ref{eq:n4}) in Appendix~B]. If the hopping parameter is higher than the other boundary value $t_{b2}>t_{b1}$, which is determined by the condition
\begin{eqnarray}
3J - 2U + 2\sqrt{U^{2} + 27t_{b2}^{2}}\,\cos\left(\phi_{U,\,t_{b2}}\right) = 0,
\label{eq:tb2}
\end{eqnarray}
the macroscopically degenerate chiral phase $\widetilde{\rm S}_{3}$ can be identified between S$_{2}$ and S$_{4}$. The phases S$_{2}$, S$_{4}$ gradually enlarge, while $\widetilde{\rm S}_{3}$ is shifted towards higher magnetic fields with the increasing $t$. As a result a vertical boundary arises between the phases S$_{2}$, S$_{4}$ at the magnetic fields $H \leq J - \frac{2U}{3} + 2t +\! \sqrt{(U+2t)^{2}+32t^{2}} - \frac{4}{3}\sqrt{U^{2} \!+\! 27t^{2}}\cos\left(\phi_{U,\,t}\right)$. The phase transition S$_{2}$--S$_{4}$ can be observed only for hopping parameters which are higher than the boundary value $t_{b3}>t_{b2}$ given by equation
\begin{eqnarray}
3J - 2U + 6t_{b3} + 3\sqrt{(U+2t_{b3})^{2}+32t_{b3}^{2}}
 {}  \hspace{1.15cm}
\nonumber\\
 {}  - 4\sqrt{U^{2} + 27t_{b3}^{2}}\,\cos\left(\phi_{U,\,t_{b3}}\right)
= 0,
\label{eq:tb3}
\end{eqnarray}
as illustrated in Fig.~\ref{fig2}(d).

It is worthy to mention that all observed ground-state boundaries are discontinuous (first-order) phase transitions, where two neighboring phases coexist together.
Their analytical expressions can be derived by comparing energies of these phases (see Appendix C). The exceptions are the phase transitions S$_{1}$--S$_{3}$ and S$_{3}$--S$_{5}$, which are depicted by dashed lines in Fig.~\ref{fig2}. At each point of these particular boundaries, spin-electron configurations of the elementary clusters corresponding to the adjacent phases S$_{1}$, S$_{3}$ (S$_{3}$, S$_{5}$) are in thermodynamic equilibrium with the third configuration ${\rm \widetilde{S}}_{2k}$ (${\rm \widetilde{S}}_{4k}$) given by the following eigenvector
and ground-state energy
\begin{eqnarray}
\label{eq:S_2,4chiral}
|{\rm \widetilde{S}}_{c}\rangle_k \!\!&=&\!\! \left\{ \begin{tabular}{l}
                                                           $|\!\downarrow\rangle_{\sigma_k}$ $\!\otimes\!$
                                                            $|n_{k}=c,S_{k}^{z}=-1\rangle_{L, R}$,\,\hspace{2mm}
                                                            $H = 0$
                                                            \\[2mm]
                                                           $|\!\uparrow\rangle_{\sigma_k}$ $\!\otimes\!$
                                                            $|n_{k}=c,S_{k}^{z}=1\rangle_{ L, R}$,\hspace{5mm} $H \geq 0$
                                                     \end{tabular}
                                               \right.\!\!,
                                                \nonumber\\
\widetilde{E}_{ck} \!\!&=&\!\! -J - \frac{3H}{2} - c\mu - t, \quad c=\{2,4\}.
\end{eqnarray}
Similarly as ${\rm \widetilde{S}}_{3}$, not yet observed configurations~(\ref{eq:S_2,4chiral}) are macroscopically degenerate due to the chiral degrees of freedom of mobile electrons [see~(\ref{eq:n2ch-}), (\ref{eq:n2ch+}), (\ref{eq:n4ch-}), (\ref{eq:n4ch+}) in Appendix~B].

For completeness, we should remark that other variants of the ground-state phase diagrams could also be discussed. However, all possible spin-electron arrangements resulting from the mutual competition between the parameters $J, t, U, \mu$, $H$ have already been presented in Fig.~\ref{fig2}, therefore we omit them in this paper.

\subsection{Electron density}
\label{subsec:3B}

\begin{figure}[t!]
\vspace{0.25cm}
    \includegraphics[width = 1.0\columnwidth]{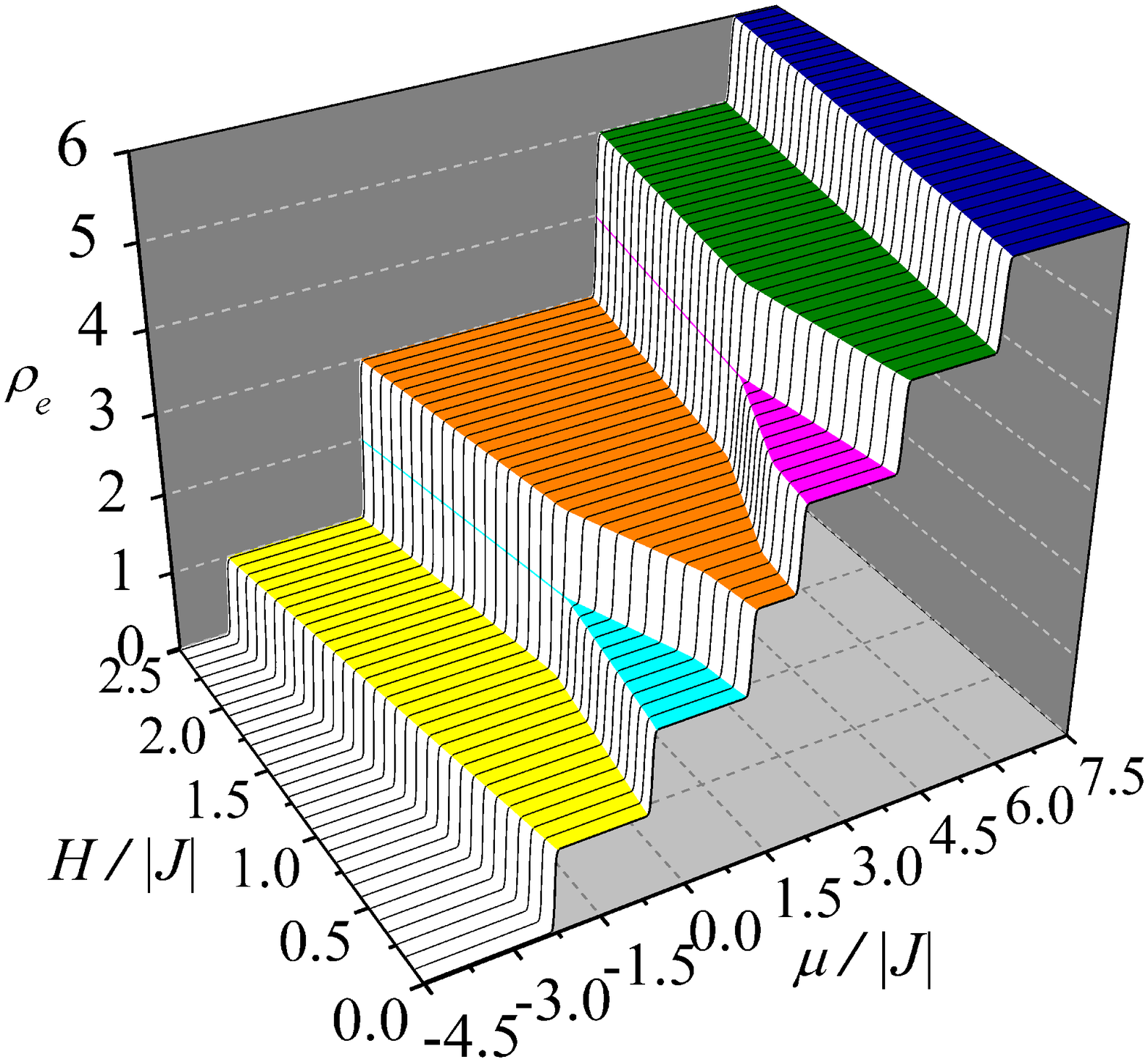}
    \\[2mm]
     \includegraphics[width = 1.0\columnwidth]{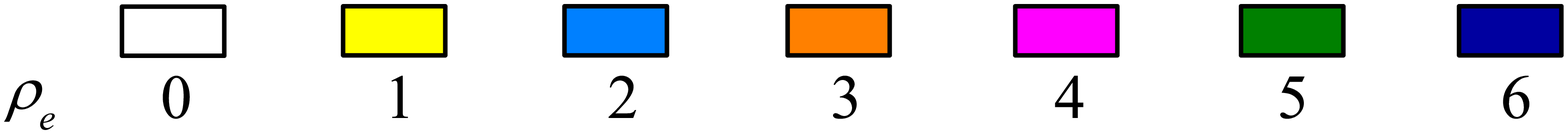}
\caption{\small (Color online) The electron density $\rho_e$ per triangular plaquette as a function of the chemical potential $\mu$ and the magnetic field $H$ for the Coulomb term $U/|J| = 3$ and the hopping parameter $t/|J| = 0.9$ at the temperature \mbox{$k_{\rm B}T/|J|=0.015$.}}
  \label{fig3}
\end{figure}

The electron content in triangular plaquettes observed within individual ground-state phases can be independently verified by the electron density per triangular plaquette $\rho_e = \left\langle \hat{n}_k\right\rangle_{\rm gcn.}$ at low enough temperatures. This physical quantity can be obtained by using the relation:
\begin{eqnarray}
\label{eq:ro}
\rho_{e} = -\left(\frac{\partial \Omega}{\partial\mu}\right)_{\!T}.
\end{eqnarray}
Typical three-dimensional (3D) plot of the electron density per triangular plaquette against the chemical potential $\mu$ and the magnetic field $H$ is presented in Fig.~\ref{fig3}. To compare the displayed results with the ground-state analysis in Subsec.~\ref{subsec:3A}, the reduced parameters $U/|J|$ and $t/|J|$ are fixed to the same values as in Fig.~\ref{fig2}(c). The temperature takes the lowest possible value for numerical calculations $k_{\rm B}T/|J|=0.015$. It is clear from Fig.~\ref{fig3} that the displayed electron density exhibits in total seven different plateaux at $\rho_e = 0, 1, 2, 3, 4, 5$ and $6$. Referring to the phase diagrams in Fig.~\ref{fig2}(c) one can conclude that the observed zero and integer plateaux reflect the electron content per plaquette in the ground-state phases S$_0$, S$_1$, S$_2$, S$_3$, ${\rm \widetilde{S}}_3$, S$_4$, S$_5$ and S$_6$, respectively. Steep continuous steps between different plateaux quite accurately reproduce real discontinuous jumps which exist at relevant first-order phase transitions at the absolute zero temperature. In accordance with common expectations, the rising temperature smoothes the observed stepwise behavior of~$\rho_e$ until the plateaux completely disappear.

\subsection{Magnetization process}
\label{subsec:3C}

To investigate the magnetization process of the spin-electron double-tetrahedral chain it is useful to introduce the single-site magnetization corresponding to the localized Ising spins $m_{I} = \langle\sigma_{k}^{z}\rangle_{\rm gcn.}$ and the magnetization of mobile electrons per triangular plaquette $m_{e} = \langle\hat{S}_{k}^{z}\rangle_{\rm gcn.}$. Both the magnetization can be obtained as derivatives of the grand free energy~(\ref{eq:Omega}) with respect to the particular magnetic fields:
\begin{eqnarray}
\label{eq:mImegcn}
m_{I} = -\left(\frac{\partial \Omega}{\partial H_{I}}\right)_{\!T},
\qquad
m_{e} = -\left(\frac{\partial \Omega}{\partial H_{e}}\right)_{\!T}.
\end{eqnarray}
In view of this notation, the total magnetization per elementary cluster $m$ can be defined as a sum of both magnetization:
\begin{eqnarray}
\label{eq:m}
m = m_{I} + m_{e}.
\end{eqnarray}
It is worth to note that the value of~(\ref{eq:m}) depends on the spin states of magnetic particles forming the elementary cluster as well as the electron density per triangular plaquette. As a consequence, the saturation magnetization per elementary cluster $m_{sat}$ can be expressed as follows:
\begin{eqnarray}
\label{eq:msat}
m_{sat} =
\begin{cases}
\dfrac{1+ \rho_e}{2} &\,\,{\rm for}\,\,\, 0 \leq \rho_e \leq 3\,,
\\[2mm]
\dfrac{7 - \rho_e}{2} &\,\,{\rm for}\,\,\, 3 < \rho_e \leq 6\,.
\end{cases}
\end{eqnarray}
\begin{figure}[t!]
\vspace{0.25cm}
    \includegraphics[width = 1.0\columnwidth]{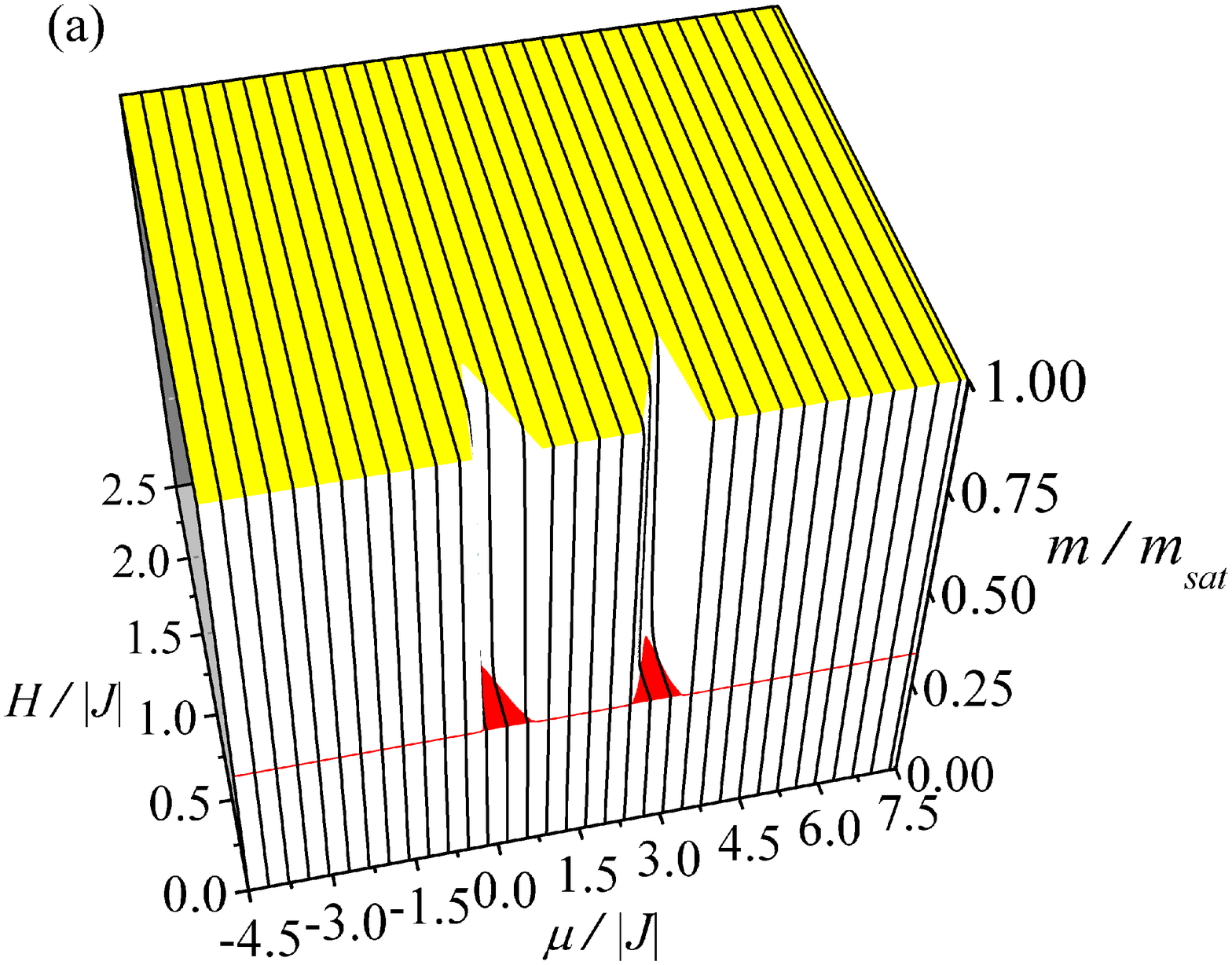}
     \\[0.3cm]
    \includegraphics[width = 1.0\columnwidth]{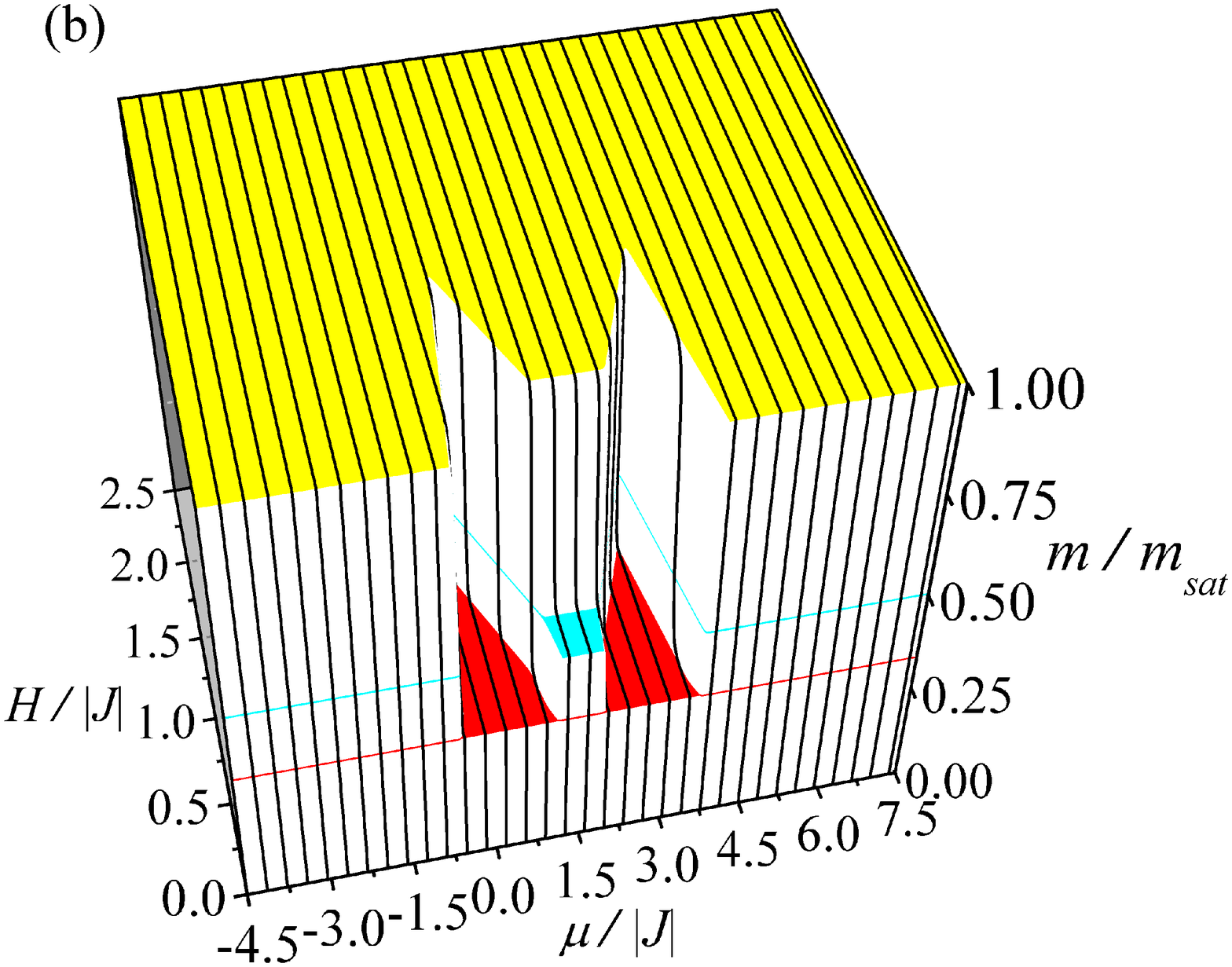}
    \\[0.3cm]
    \includegraphics[width = 1.0\columnwidth]{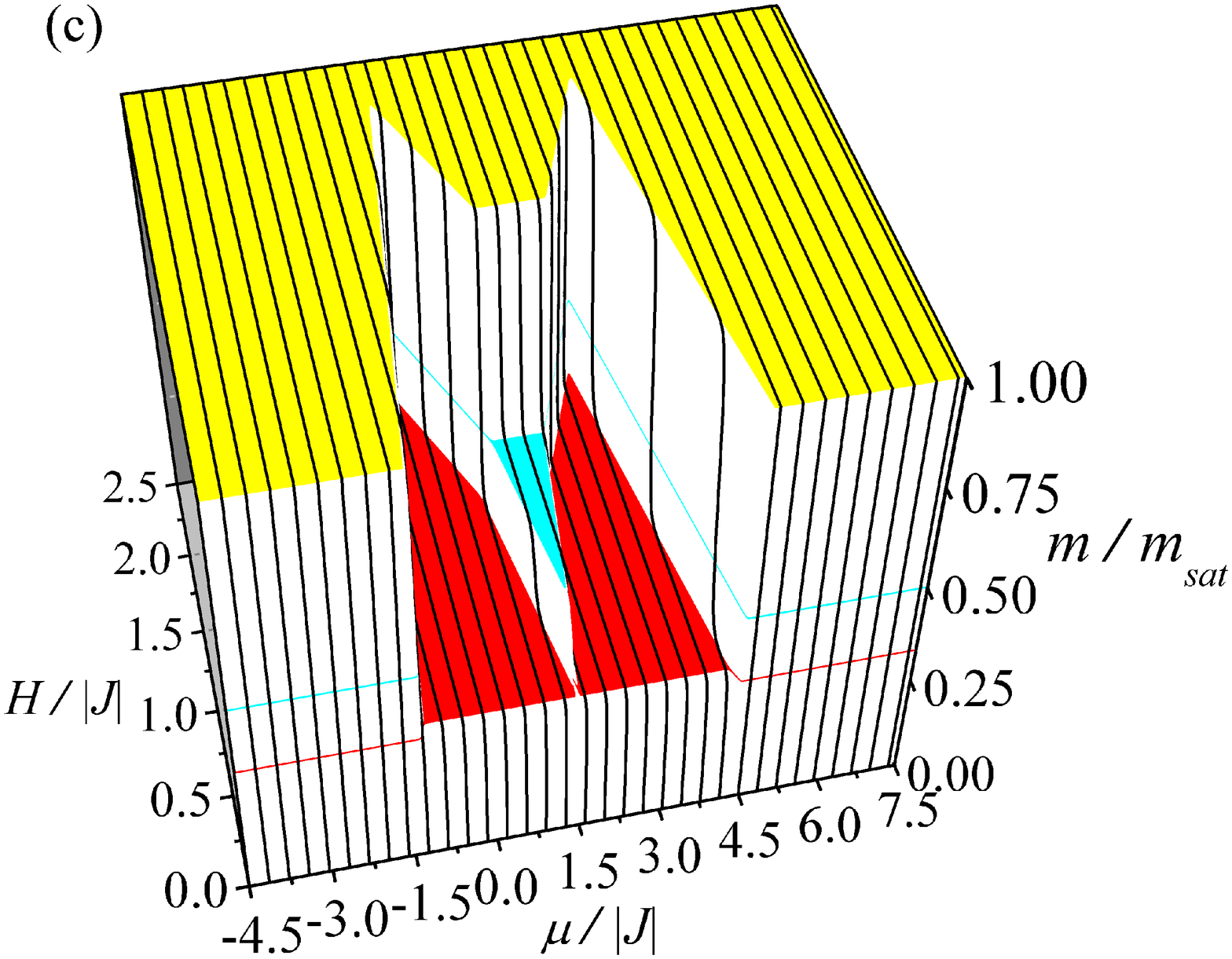}
    \\[2mm]
     \includegraphics[width = 1.0\columnwidth]{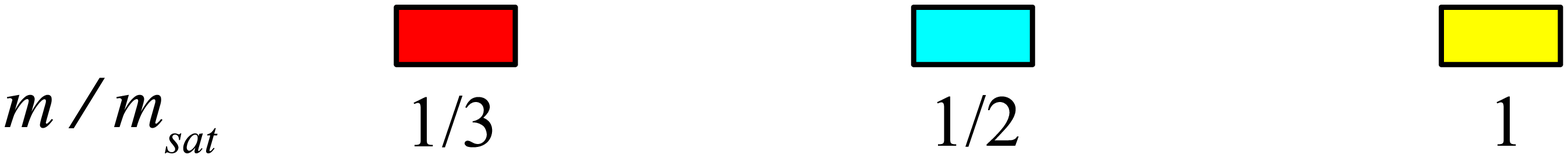}
\caption{\small (Color online) 3D views of the magnetization process for the Coulomb term $U/|J| = 3$ and the hopping parameters (a)~$t/|J| = 0.7$, (b)~$t/|J| = 0.9$, (c)~$t/|J| = 1.3$ at the temperature $k_{\rm B}T/|J| = 0.015$.}
  \label{fig4}
\vspace{-1.2cm}
\end{figure}

Figure~\ref{fig4} presents 3D views of the magnetization process for the fixed Coulomb term $U/|J| = 3$ and three representative values of the hopping parameter $t/|J| = 0.7, 0.9, 1.3$ at the temperature $k_{\rm B}T/|J| = 0.015$. Obviously, besides a direct steep increase from the zero to saturation value in the zero-field limit, 3D magnetization curves may include one or two fractional plateaux at the one-third and one-half of the saturation magnetization before reaching the saturation value. Referring to Fig.~\ref{fig2} one can conclude that the region where the total magnetization reaches the saturation value corresponds to the phases S$_{0}$, S$_{1}$, S$_{3}$, S$_{5}$ and S$_{6}$. On the other hand, the one-third plateau reflects the presence of the phases S$_{2}$ and S$_{4}$, while the one-half plateau corresponds to the chiral phase $\widetilde{\rm S}_{3}$.
The analytical expressions of individual plateaux depend on the total spin~$S^{z}_{k}$ and the electron density $\rho_e$ of triangular plaquettes:
\begin{eqnarray}
\label{eq:m/msat}
\frac{m}{m_{sat}} =
\begin{cases}
\dfrac{1+2S^{z}_{k}}{1+ \rho_e} &\quad {\rm for}\,\,\, \rho_e = 2\,\,{\rm and}\,\,3,
\\[4mm]
\dfrac{1+2S^{z}_{k}}{7 - \rho_e} &\quad {\rm for}\,\,\, \rho_e = 4.
\end{cases}
\end{eqnarray}
For completeness, we note that rational magnetization plateaux and steep steps displayed in Fig.~\ref{fig3} are not real due to the finite temperature, but they very precisely reproduce real plateaux and discontinuous jumps appearing in the zero-temperature magnetization curves.

\section{Gradual electron doping}
\label{sec:4}

In this section, we will examine ground-state properties and the low-temperature magnetization process upon a gradual (continuous) electron doping. The electron density in the model is continuously increasing from the zero up to the maximum value $6N$ during this mechanism. Therefore, it is necessary to pass from the grand canonical ensemble, in which the electron density and other physical quantities are controlled by varying the chemical potential, into the canonical one with the inverse dependence. In the context of thermodynamics, the inversion $\rho_{e}(\mu)\to\mu(\rho_{e})$ can be performed via the Legendre transformation~\cite{Shell14}:
\begin{eqnarray}
\label{eq:legendre}
G(\rho_e, \ldots) = \Omega\left[\mu(\rho_e), \ldots\right] + \mu(\rho_e)\rho_e.
\end{eqnarray}
Obviously, the above relation connects the Gibbs free energy $G$ with the grand free energy $\Omega$ that are functions of the conjugate variables $\rho_e$ and $\mu$,  respectively.
The analytical expression for the Gibbs free energy~(\ref{eq:legendre}) can be further utilized for a calculation of the sublattice magnetization $m_I = \langle\sigma_{k}^{z}\rangle_{\rm cn.}$ and $m_e = \langle\hat{S}_{k}^{z}\rangle_{\rm cn.}$:  \begin{eqnarray}
\label{eq:mImecn}
m_{I} = -\left(\frac{\partial G}{\partial H_{I}}\right)_{\!T},
\qquad
m_{e} = -\left(\frac{\partial G}{\partial H_{e}}\right)_{\!T},
\end{eqnarray}
as well as some fundamental correlation functions:
\begin{eqnarray}
\label{eq:Be}
B_{e}\!\!&=&\!\! \Big\langle\!\sum_{\gamma \in\{\uparrow, \downarrow\}}\!\sum_{j=1}^3 \!\big(\hat{c}_{kj,\gamma}^{\dag}\hat{c}_{k(j+1){_{\rm mod 3}},\gamma} \!+ {\rm H.c.}\big)\!\Big\rangle_{\rm cn.}
\nonumber\\
\!\!&=&\!\!
-\left(\frac{\partial G}{\partial t}\right)_{\!T},
\\
\label{eq:CIe}
C_{Ie}\!\!&=&\!\! \big\langle\sigma_{k}^{z}\hat{S}_{k}^{z}\big\rangle_{\rm cn.} = \frac{1}{2}\!\left(\frac{\partial G}{\partial J}\right)_{\!T},
\\
\label{eq:De}
D_{e}\!\!&=&\!\! \Big\langle\sum_{j = 1}^{3}\hat{n}_{kj,\uparrow}\hat{n}_{kj,\downarrow}\Big\rangle_{\rm cn.} = \left(\frac{\partial G}{\partial U}\right)_{\!T}.
\end{eqnarray}
The function $B_{e}$ has no physical meaning, while $C_{Ie}$ describes the correlation between electrons in the triangular plaquette and their nearest Ising neighbor and $D_{e}$ reflects a double occupancy of plaquette sites.

It should be stressed that the sublattice magnetization~(\ref{eq:mImecn}) and the correlation functions~(\ref{eq:Be})--(\ref{eq:De}) allow one to reliably describe an evolution of the possible ground-state configurations and the magnetization process upon continuous change of the electron content in the whole parameter space of the model~(\ref{eq:H}). In spite of this generality, we restrict our following analysis in the same manner as in Sec.~\ref{sec:3}, i.e. to the ferromagnetic Ising interaction $J<0$, the uniform magnetic field $H=H_I=H_e$ and the fixed Coulomb term $U/|J|=3$.

\subsection{Ground state}
\label{subsec:4A}

In the limit of zero temperature one may expect several regions with varying the particle density upon the continuous electron doping due to a competition between model parameters and the external magnetic field. To characterize these regions, we have solved numerically the set of Eqs.~(\ref{eq:mImecn})--(\ref{eq:De}) and examined the trends of each quantity depending on $J$, $t$, $U$, $H$ and $\rho_{e}$.

At the zero magnetic field, the Ising spins located at nodal lattice sites are either frustrated or may occupy one of two possible spin states $\sigma_k=-1/2$ or $1/2$ depending on whether the total spin of the neighboring triangular plaquette is zero or not, respectively. As a result, the sublattice magnetization $m_{i}$ is zero in the whole zero-field space. The sublattice magnetization $m_{e}$ and the correlation function $C_{Ie}$ are also zero. On the other hand, for any $H>0$, the Ising spins are polarized into the field direction and the sublattice magnetization corresponding to these particles reaches the saturation value $m_{i}=1/2$. Other quantities $m_{e}$, $B_{e}$, $C_{Ie}$, $D_{e}$ are functions of the electron density $\rho_e$. A careful analysis reveals that the zero-temperature parameter plane $\rho_{e}- H$ contains in total eleven different regions:
\vspace{1mm}
\newline
$\bullet$ \underline{Region ${\rm R}_{0\to1}$} (for $0<\rho_e<1$):
\begin{eqnarray}
\label{eq:R01}
\begin{tabular}{r@{\,\,=}l p{22mm} r@{\,\,=}l  p{15mm} r@{\,\,=}l p{10mm}}
$m_{e}$&&$\dfrac{\rho_{e}}{2}$, & $C_{Ie}$&& $\dfrac{\rho_{e}}{4}$, \\[2mm]
$B_{e}$&&$2\rho_{e}$, & $D_{e}$&&$0$;
\end{tabular}
\end{eqnarray}
In this region, the sublattice magnetization $m_{e}$, the correlation functions $C_{Ie}$ and the quantity $B_{e}$ start from zero and linearly increase with $\rho_e$, because empty sites of individual non-magnetic plaquettes are gradually occupied by single electrons with spins polarized into the magnetic-field direction. This particle distribution is confirmed by the zero value of the double occupancy~$D_{e}$.
\vspace{1mm}
\newline
$\bullet$ \underline{Region ${\rm R}_{1\to\widetilde{2}\to3}$} (for $1<\rho_e<3$):
\begin{eqnarray}
\label{eq:R12ch3}
\begin{tabular}{r@{\,\,=}l p{22mm} r@{\,\,=}l  p{15mm} r@{\,\,=}l p{10mm}}
$m_{e}$&&$\dfrac{\rho_{e}}{2}$, & $C_{Ie}$&& $\dfrac{\rho_{e}}{4}$, \\[2mm]
$B_{e}$&&$3 - \rho_{e}$, & $D_{e}$&&$ 0$;
\end{tabular}
\end{eqnarray}
In this region, mobile electrons with spins oriented in the field direction continue in the occupation of empty plaquette sites. As a consequence, $m_{e}$, $C_{Ie}$ have the same analytical expressions as in R$_{0\to1}$. The double occupancy $D_{e}$ also remains zero, while $B_{e}$ linearly decreases with the increasing~$\rho_e$. The electron distribution upon doping process is as follows: For $1<\rho_e<2$, the number of triangular plaquettes occupied by two electrons increases at the expense of those that are occupied by a single particle. Referring to the ground-state analysis in Sec.~\ref{sec:3}, it can be concluded that emerging one-third-filled plaquettes exhibit the non-zero chirality. If $2<\rho_e<3$, then the number of chiral plaquettes decreases and the half-filled plaquettes with all particles polarized into the magnetic-field direction are created with the increasing~$\rho_e$.
\vspace{1mm}
\newline
$\bullet$ \underline{Region ${\rm R}_{3\to\widetilde{4}\to5}$} (for $3<\rho_e<5$):
\begin{eqnarray}
\label{eq:R34ch5}
\begin{tabular}{r@{\,\,=}l p{22mm} r@{\,\,=}l  p{15mm} r@{\,\,=}l p{10mm}}
$m_{e}$&&$3-\dfrac{\rho_{e}}{2}$, & $C_{Ie}$&& $\dfrac{3}{2}-\dfrac{\rho_{e}}{4}$, \\[2mm]
$B_{e}$&&$\rho_{e} - 3$, & $D_{e}$&&$\rho_{e} - 3$;
\end{tabular}
\end{eqnarray}
The distribution of mobile electrons within this region upon doping process is as follows: Because there is no empty plaquette site, triangular plaquettes with two singly occupied and one doubly occupied sites are occurred for $3<\rho_e<4$. According to the ground-state analysis in Sec.~\ref{sec:3}, created plaquettes have the non-zero chirality.  For $4<\rho_e<5$, the electrons continue in double occupation of plaquette sites upon doping. This leads to the creation of non-chiral plaquettes with one singly occupied and two doubly occupied sites at the expense of chiral ones. As a result, $D_e$ linearly increases, while $m_e$ and $C_{Ie}$ decrease with the increasing~$\rho_e$. The quantity $B_e$ has the same analytical expression as $D_e$.
\vspace{1mm}
\newline
$\bullet$ \underline{Region ${\rm R}_{5\to6}$} (for $5<\rho_e<6$):
\begin{eqnarray}
\label{eq:R56}
\begin{tabular}{r@{\,\,=}l p{22mm} r@{\,\,=}l  p{15mm} r@{\,\,=}l p{10mm}}
$m_{e}$&&$3-\dfrac{\rho_{e}}{2}$, & $C_{Ie}$&& $\dfrac{3}{2}-\dfrac{\rho_{e}}{4}$, \\[2mm]
$B_{e}$&&$12 - 2\rho_{e}$, & $D_{e}$&&$\rho_{e} - 3$;
\end{tabular}
\end{eqnarray}
This region appears only for sufficiently great electron densities. The electrons continue in occupation of singly occupied plaquette sites. As a consequence, the fully-filled plaquettes emerge upon electron doping. The quantities $m_{e}$, $C_{Ie}$, $D_e$ have the same analytical expressions as in R$_{3\to\widetilde{4}\to5}$, while $B_{e}$ linearly decreases upon increasing~$\rho_e$.
\vspace{1mm}
\newline
$\bullet$ \underline{Region ${\rm R}_{1\to2}$} (for $1<\rho_e<2$):
\begin{eqnarray}
\label{eq:R12}
\begin{tabular}{r@{\,\,=}l p{22mm} r@{\,\,=}l  p{15mm} r@{\,\,=}l p{10mm}}
$m_{e}$&&$1-\dfrac{\rho_{e}}{2}$, & $C_{Ie}$&& $\dfrac{1}{2}-\dfrac{\rho_{e}}{4}$, \\[2mm]
\multicolumn{6}{l}{$B_{e}=2 - \left[1 - \dfrac{U+18t}{\sqrt{(U+2t)^2+32t^2}}\right]\!(\rho_e-1)$,}  \\[4mm]
\multicolumn{6}{l}{$D_{e}=\left[\dfrac{1}{2} - \dfrac{U+2t}{2\sqrt{(U+2t)^2+32t^2}}\right]\!(\rho_e-1)$;}
\end{tabular}
\end{eqnarray}
In this region, an interplay between the parameters $U$, $t$ favors the electron distribution with opposite spins. As a result, the plaquettes occupied by electron pairs in a quantum superposition of six intrinsic antiferromagnetic and three non-magnetic states are formed upon doping process. At the same time, one-sixth-filled plaquettes with the total spin $S_k^z=1/2$ disappear. Since the total spin of the created plaquettes is zero, the magnetization $m_e$ and the correlation function $C_{ie}$ linearly decrease upon doping process. Other two quantities $B_{e}$, $D_{e}$ linearly increase with $\rho_e$, but their analytical expressions are more complex, since their values depend on $U$ and $t$.
\vspace{1mm}
\newline
$\bullet$ \underline{Region ${\rm R}_{2\to3}$} (for $2<\rho_e<3$):
\begin{eqnarray}
\label{eq:R23}
\begin{tabular}{r@{\,\,=}l p{22mm} r@{\,\,=}l  p{15mm} r@{\,\,=}l p{10mm}}
$m_{e}$&&$\dfrac{3\rho_{e}}{2}-3$, & $C_{Ie}$&& $\dfrac{3\rho_{e}}{4}-\dfrac{3}{2}$, \\[3mm]
\multicolumn{6}{l}{$B_{e}= \left[1+\dfrac{U+18t}{\sqrt{(U+2t)^2+32t^2}}\right]\!(3-\rho_e)$,}  \\[4mm]
\multicolumn{6}{l}{$D_{e}=\left[\dfrac{1}{2} - \dfrac{U+2t}{2\sqrt{(U+2t)^2+32t^2}}\right]\!(3-\rho_e)$;}
\end{tabular}
\end{eqnarray}
This region is a continuation of ${\rm R}_{1\to2}$ when the electron density exceeds the value $\rho_e=2$. The magnetization $m_e$ and the correlation function $C_{ie}$ linearly increase up to their maximum values $m_e=3/2$ and $C_{ie}=3/4$, while $B_{e}$ and $D_{e}$ decrease to the zero with the increasing $\rho_e$. It can be thus concluded that all the particles in triangular plaquettes prefer the seating at different sites, since they are fully polarized into the field direction. The intrinsic antiferromagnetic and non-magnetic states disappear.
\vspace{1mm}
\newline
$\bullet$ \underline{Region ${\rm R}_{3\to4}$} (for $3<\rho_e<4$):
\begin{eqnarray}
\label{eq:R34}
\begin{tabular}{r@{\,\,=}l p{22mm} r@{\,\,=}l  p{15mm} r@{\,\,=}l p{10mm}}
$m_{e}$&&$6-\dfrac{3\rho_{e}}{2}$, & $C_{Ie}$&& $3-\dfrac{3\rho_{e}}{4}$, \\[2mm]
\multicolumn{6}{l}{$B_{e}= \left[1+\dfrac{U+18t}{\sqrt{(U+2t)^2+32t^2}}\right]\!(\rho_e-3)$,}  \\[4mm]
\multicolumn{6}{l}{$D_{e}=\left[\dfrac{3}{2} - \dfrac{U+2t}{2\sqrt{(U+2t)^2+32t^2}}\right]\!(\rho_e-3)$;}
\end{tabular}
\end{eqnarray}
In this region, there is no empty plaquette site in the system, therefore two-thirds-filled plaquettes involving doubly occupied sites start to appear upon electron doping. However, the minimization of the energy requires a quantum superposition of the intrinsic antiferromagnetic  and/or non-magnetic states of mobile electrons. The total spin of emerging plaquettes is thus zero. This scenario is well described by the decreasing magnetization $m_e$ and correlation function $C_{Ie}$. The double occupancy $D_e$ and the quantity $B_e$ are increasing functions of $\rho_e$.
\vspace{1mm}
\newline
$\bullet$ \underline{Region ${\rm R}_{4\to5}$} (for $4<\rho_e<5$):
\begin{eqnarray}
\label{eq:R45}
\begin{tabular}{r@{\,\,=}l p{22mm} r@{\,\,=}l  p{15mm} r@{\,\,=}l p{10mm}}
$m_{e}$&&$\dfrac{\rho_{e}}{2}-2$, & $C_{Ie}$&& $\dfrac{\rho_{e}}{4}-1$, \\[2mm]
\multicolumn{6}{l}{$B_{e}=2 - \left[1 - \dfrac{U+18t}{\sqrt{(U+2t)^2+32t^2}}\right]\!(5-\rho_e)$,}  \\[4mm]
\multicolumn{6}{l}{$D_{e}=2-\left[\dfrac{1}{2} + \dfrac{U+2t}{2\sqrt{(U+2t)^2+32t^2}}\right]\!(5-\rho_e)$;}
\end{tabular}
\end{eqnarray}
In this region, the interplay between the parameters $U$, $t$ and the external field $H$ favors the formation of plaquettes with the total spin $S_k^z=1/2$, in which one lattice site is occupied by a single electron polarized into the field direction, while other ones are occupied by electron pairs with opposite spins. At the same time, the number of plaquettes with the total spin $S_k^z=0$ is gradually diminished. The evolution of afore-described particle distribution is confirmed by the linear increase of $m_e$, $C_{Ie}$ and also $D_e$. The value of the quantity $B_e$ decreases upon the electron doping.
\vspace{1mm}
\newline
$\bullet$ \underline{Region ${\rm R}_{2\to\widetilde{3}}$} (for $2<\rho_e<3$):
\begin{eqnarray}
\label{eq:R23ch}
\begin{tabular}{r@{\,\,=}l p{22mm} r@{\,\,=}l  p{15mm} r@{\,\,=}l p{10mm}}
$m_{e}$&&$\dfrac{\rho_{e}}{2}-1$, & $C_{Ie}$&& $\dfrac{\rho_{e}}{4}-\dfrac{1}{2}$,
\\[2mm]
\multicolumn{6}{l}{$B_{e}=\left[1+\dfrac{U+18t}{\sqrt{(U+2t)^2+32t^2}}\right]\!(3-\rho_e)$}  \\[4mm]
\multicolumn{6}{l}{\hspace{0.8cm}$+\dfrac{18t}{\sqrt{U^2+27t^2}}\cos\left(\phi_{U,\,t}\right)(\rho_e-2)$}
\\[4mm]
\multicolumn{6}{r}{$- \dfrac{2\sqrt{U^2+27t^2}}{3}\dfrac{\partial\phi_{U,\,t}}{\partial t}\sin\left(\phi_{U,\,t}\right)(\rho_e-2)$,}
\\[4mm]
\multicolumn{6}{l}{$D_{e}=\dfrac{2}{3} + \left[\dfrac{1}{6} + \dfrac{U+2t}{2\sqrt{(U+2t)^2+32t^2}}\right]\!(\rho_e-3)$}
\\[4mm]
\multicolumn{6}{l}{\hspace{0.8cm}$-\,\dfrac{2U}{3\sqrt{U^2+27t^2}}\cos\left(\phi_{U,\,t}\right)(\rho_e-2)$}
\\[4mm]
\multicolumn{6}{r}{$+\, \dfrac{2\sqrt{U^2+27t^2}}{3}\dfrac{\partial\phi_{U,\,t}}{\partial U}\sin\left(\phi_{U,\,t}\right)(\rho_e-2)$;}
\end{tabular}
\end{eqnarray}
In this region, the electron doping results in a formation of chiral half-filled plaquettes with the total spin $S_k^z=1/2$ at the expense of non-chiral ones with $S_k^z=0$. The number of electrons with spins oriented in the magnetic-field direction and doubly-occupied plaquette sites disappear during this process. The creation of chiral plaquettes leads to the increase of $m_e$, $C_{ie}$ and $D_e$. By contrast, the value of $B_e$ decreases if the electron density continuously increases.
\vspace{1mm}
\newline
$\bullet$ \underline{Region ${\rm R}_{\widetilde{3}\to4}$} (for $3<\rho_e<4$):
\begin{eqnarray}
\label{eq:R3ch4}
\begin{tabular}{r@{\,\,=}l p{22mm} r@{\,\,=}l  p{15mm} r@{\,\,=}l p{10mm}}
$m_{e}$&&$2-\dfrac{\rho_{e}}{2}$, & $C_{Ie}$&& $1-\dfrac{\rho_{e}}{4}$,
\\[2mm]
\multicolumn{6}{l}{$B_{e}=\left[1+\dfrac{U+18t}{\sqrt{(U+2t)^2+32t^2}}\right]\!(\rho_e-3)$}  \\[4mm]
\multicolumn{6}{l}{\hspace{1.25cm}$+\dfrac{18t}{\sqrt{U^2+27t^2}}\cos\left(\phi_{U,\,t}\right)(4-\rho_e)$}
\\[4mm]
\multicolumn{6}{r}{$- \dfrac{2\sqrt{U^2+27t^2}}{3}\dfrac{\partial\phi_{U,\,t}}{\partial t}\sin\left(\phi_{U,\,t}\right)(4-\rho_e)$,}
\\[4mm]
\multicolumn{6}{l}{$D_{e}=\dfrac{2}{3} + \left[\dfrac{5}{6} - \dfrac{U+2t}{2\sqrt{(U+2t)^2+32t^2}}\right]\!(\rho_e-3)$}
\\[4mm]
\multicolumn{6}{l}{\hspace{1.25cm}$+\,\dfrac{2U}{3\sqrt{U^2+27t^2}}\cos\left(\phi_{U,\,t}\right)(\rho_e-4)$}
\\[4mm]
\multicolumn{6}{r}{$-\, \dfrac{2\sqrt{U^2+27t^2}}{3}\dfrac{\partial\phi_{U,\,t}}{\partial U}\sin\left(\phi_{U,\,t}\right)(\rho_e-4)$;}
\end{tabular}
\end{eqnarray}
The energy minimization of this region requires a formation of non-magnetic and intrinsic antiferromagnetic states of four electrons per plaquette during the electron doping. As a result, the number of non-chiral two-thirds-filled plaquettes with the zero total spin  gradually increases at the expense of chiral ones which are occupied by three mobile electrons. This scenario is well confirmed by the decrease of $m_e$, $C_{Ie}$, $B_e$ and the increase of $D_e$ as $\rho_e$ increases.
\vspace{1mm}
\newline
$\bullet$ \underline{Region ${\rm R}_{2\to4}$} (for $2<\rho_e<4$):
\begin{eqnarray}
\label{eq:R24}
\begin{tabular}{r@{\,\,=}l p{22mm} r@{\,\,=}l  p{15mm} r@{\,\,=}l p{10mm}}
$m_{e}$&&$0$, & $C_{Ie}$&& $0$, \\[2mm]
\multicolumn{6}{l}{$B_{e}=1 + \dfrac{U+18t}{\sqrt{(U+2t)^2+32t^2}}$,}  \\[4mm]
\multicolumn{6}{l}{$D_{e}=\dfrac{\rho_e-1}{2} - \dfrac{U+2t}{2\sqrt{(U+2t)^2+32t^2}}$;}
\end{tabular}
\end{eqnarray}
This region is the only one, where triangular plaquettes are gradually doped by electron pairs and total spins of plaquettes are conserved to the zero value. The reason is a creation of intrinsic antiferromagnetic and non-magnetic states of mobile electrons during the doping process. This evolution of electron distribution is confirmed by the zero magnetization $m_e$ and the zero correlation function $C_{ie}$. The quantity $B_e$ also remains constant upon the increasing $\rho_e$, but its value is given by the parameters $U$, $t$. On the other hand, the quantity $D_e$ linearly increases with $\rho_e$ due to the increasing number of doubly occupied plaquette sites.
\begin{figure*}[t!]
    \includegraphics[width = 0.45\textwidth]{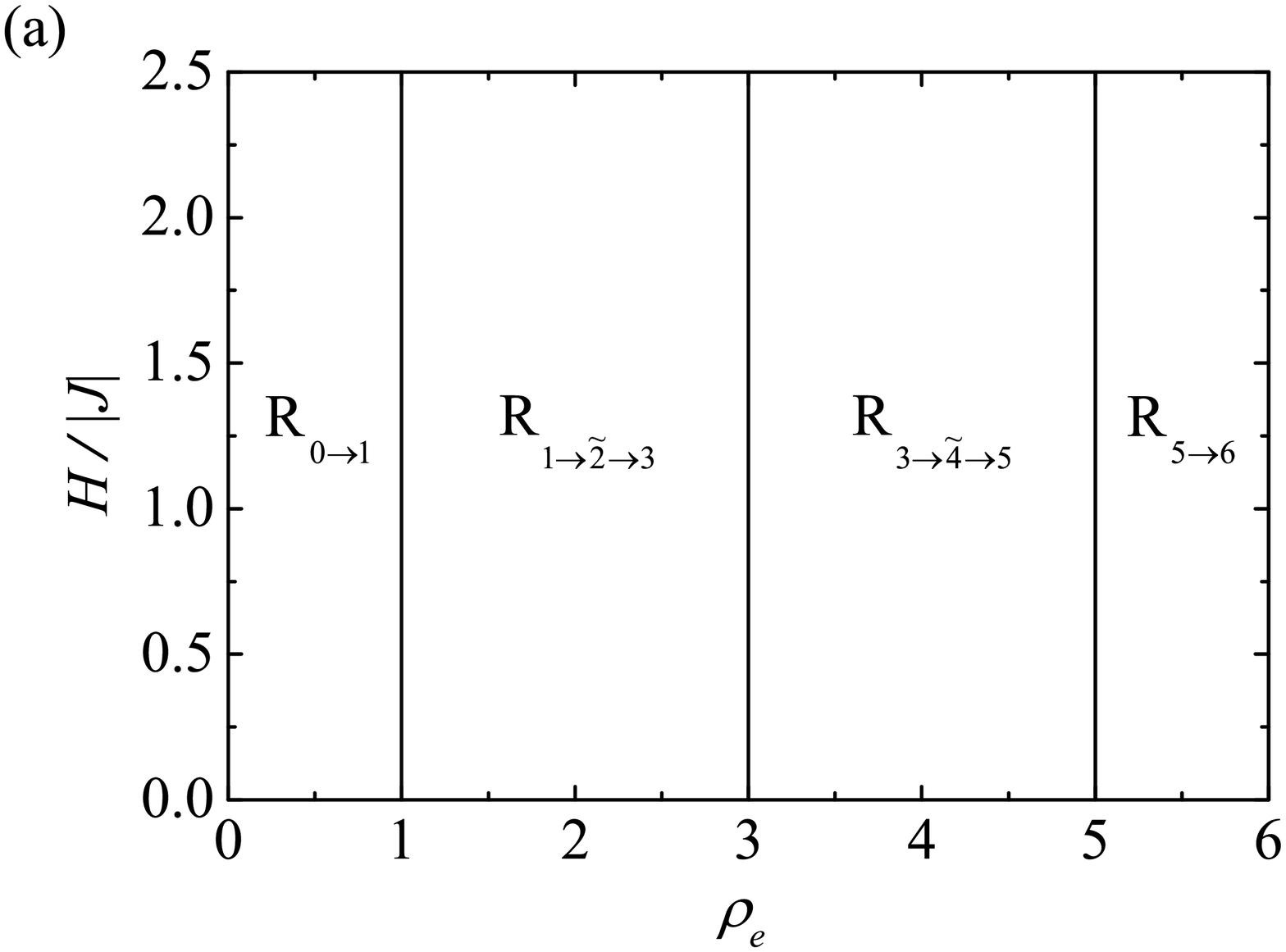}
     \hspace{0.0cm}
    \includegraphics[width = 0.45\textwidth]{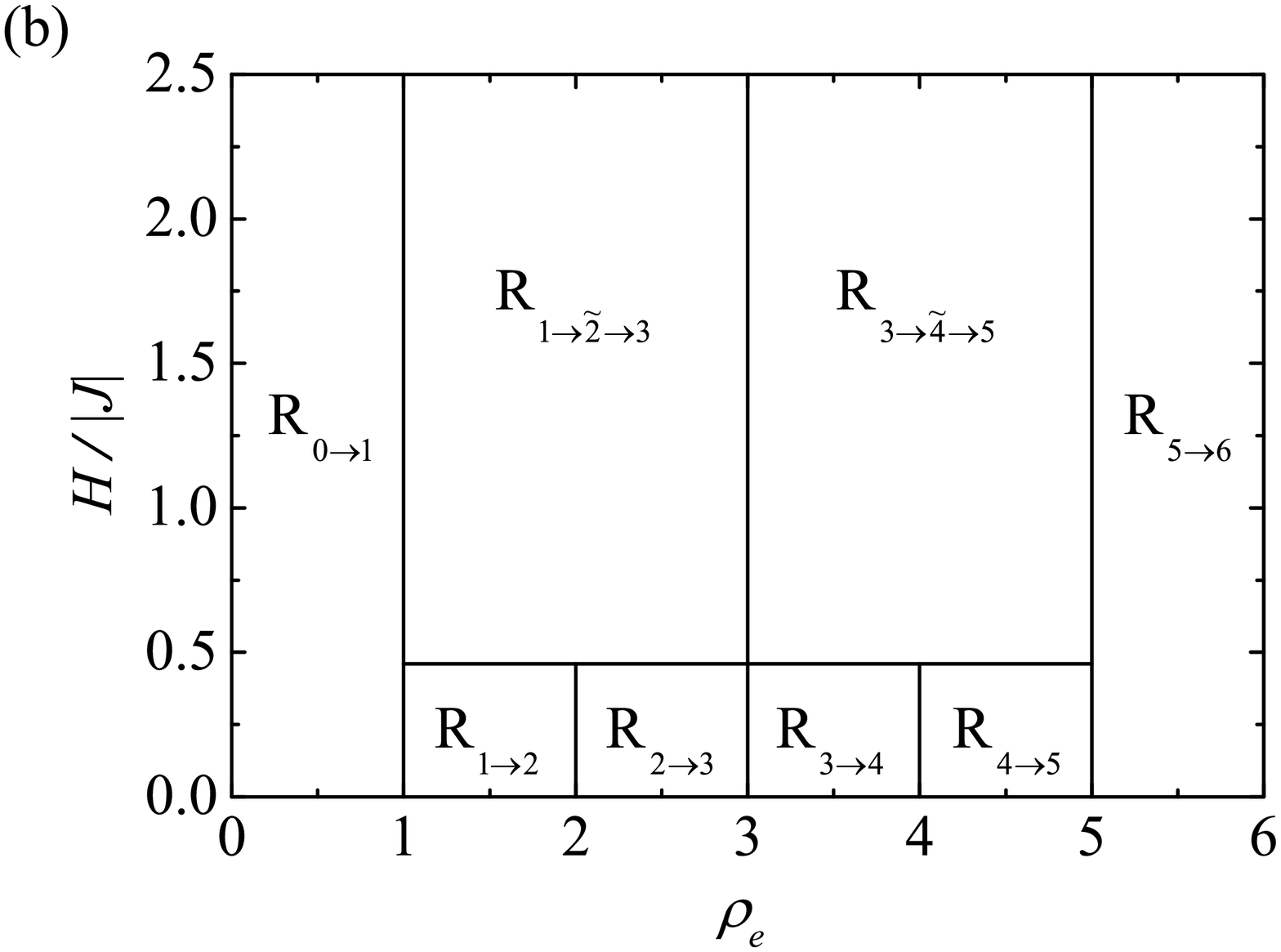}
    \\[0.25cm]
    \includegraphics[width = 0.45\textwidth]{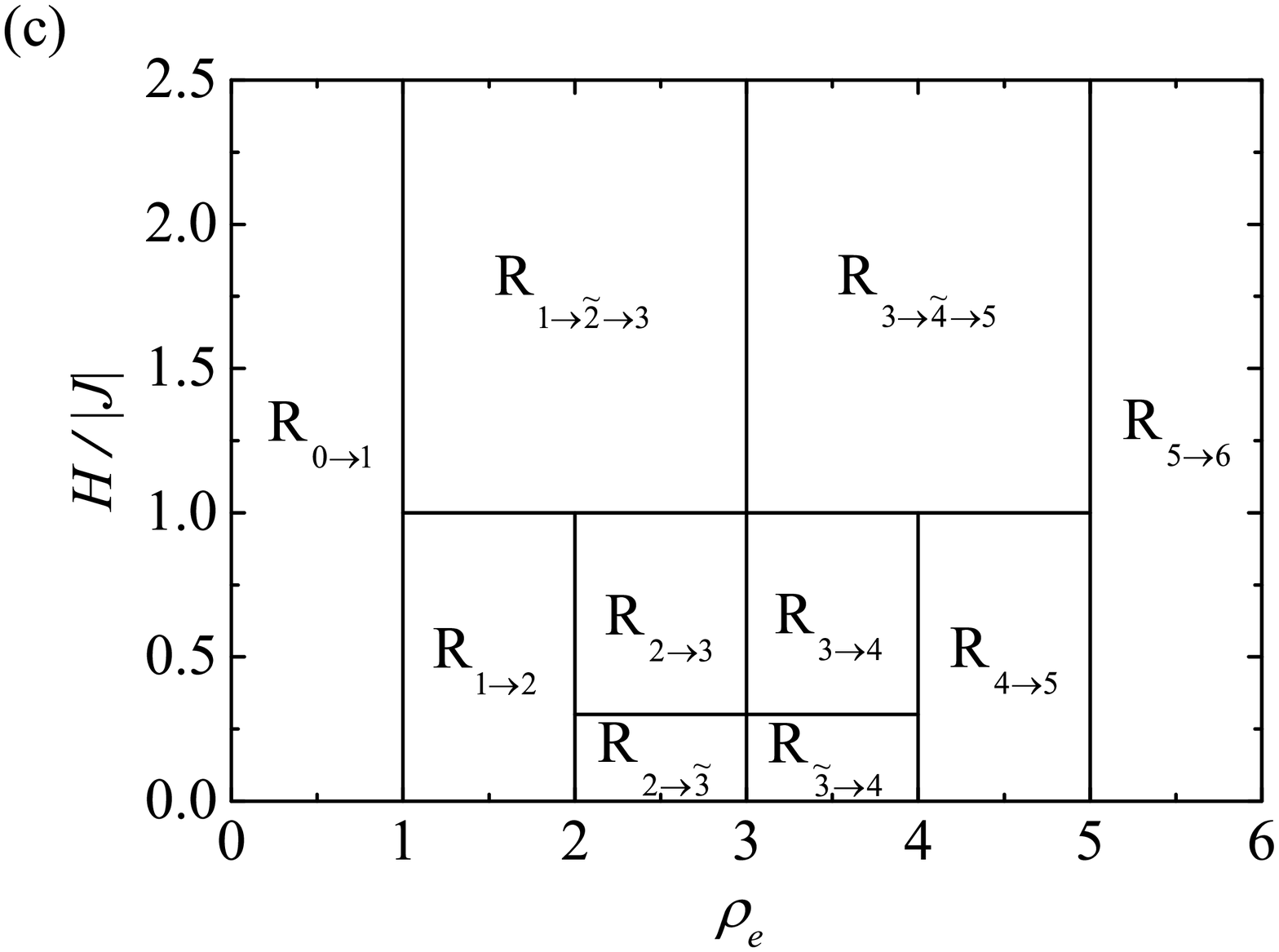}
     \hspace{0.0cm}
    \includegraphics[width = 0.45\textwidth]{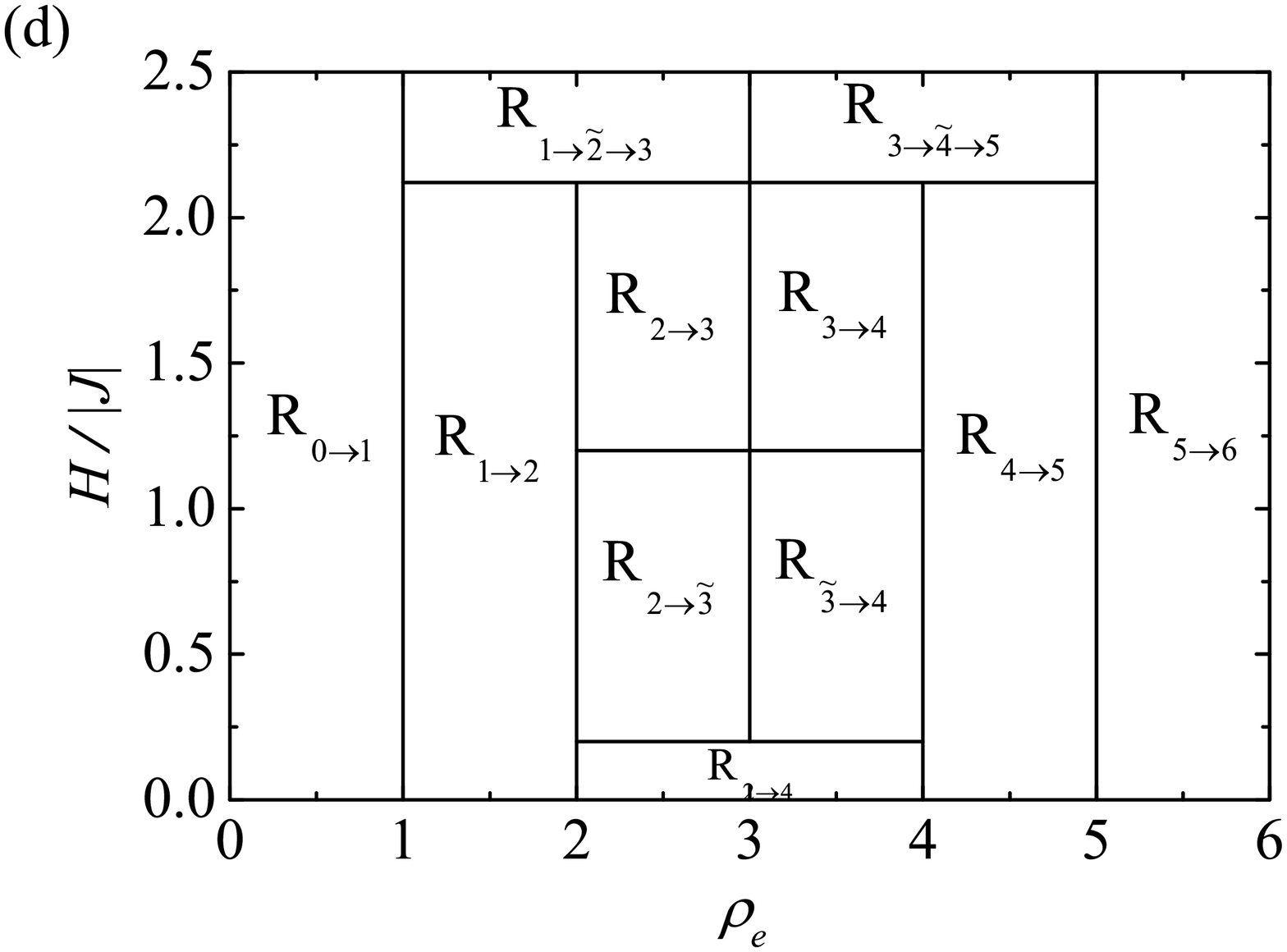}
  \caption{\small Ground-state phase diagrams of the model~(\ref{eq:H}) in the $\rho_{e}-H$ parameter plane for the fixed Coulomb term $U/|J| = 3$ and four representative values of the hopping parameter (a)~$t/|J| = 0.4$, (b)~$t/|J| = 0.7$, (c)~$t/|J| = 0.9$, (d)~$t/|J| = 1.3$.}
  \label{fig5}
\end{figure*}

In order to present locations of afore-described regions in terms of mutual interplay between the model parameters and the magnetic field, four zero-temperature phase diagrams in the parameter plane $\rho_{e}- H$ are depicted in Fig.~\ref{fig5}. It is clear from Fig.~\ref{fig5}(a) that for the kinetic parameter smaller than the value $t_{b1}$ given by Eq.~(\ref{eq:tb1}), the model passes a sequence of the phase transitions ${\rm R}_{0\to1}\to {\rm R}_{1\to\widetilde{2}\to3}\to{\rm R}_{3\to\widetilde{4}\to5}\to{\rm R}_{5\to6}$ upon a continuous increase of the electron content. If the reverse condition $t>t_{b1}$ is valid, other four regions R$_{1\to2}$, R$_{2\to3}$, R$_{3\to4}$, R$_{4\to5}$ appear in the ground state [see Figs.~\ref{fig5}(b)--(d)]. The existence of these regions is limited to the electron densities $1<\rho_e<5$ and the magnetic fields $H<J-\frac{U}{2}+\frac{1}{2}\sqrt{(U+2t)^2+32t^2}$. The last three regions R$_{2\to\widetilde{3}}$, R$_{\widetilde{3}\to4}$ and R$_{2\to4}$ can be found exclusively in the density range $2<\rho_e<4$. The first two regions are present in the $\rho_e-H$ plane at the fields $H< J - \frac{2U}{3} +\frac{2}{3}\sqrt{U^2+27t^2}\cos\left(\phi_{U,\,t}\right)$ if the hopping parameter is higher than the boundary value $t_{b2}$ given by Eq.~(\ref{eq:tb2}), while the third one occurs only for the hopping terms $t>t_{b3}$ [Eq.~(\ref{eq:tb3})] in the field region $H<J + 2t - \frac{2U}{3} + \sqrt{(U+2t)^2+32t^2} - \frac{4}{3}\sqrt{U^2+27t^2}\cos\left(\phi_{U,\,t}\right)$, as displayed in Figs.~\ref{fig5}(c), (d).

\subsection{Chemical potential}
\label{subsec:4B}

\begin{figure}[t!]
\vspace{0.25cm}
    \includegraphics[width = 1.0\columnwidth]{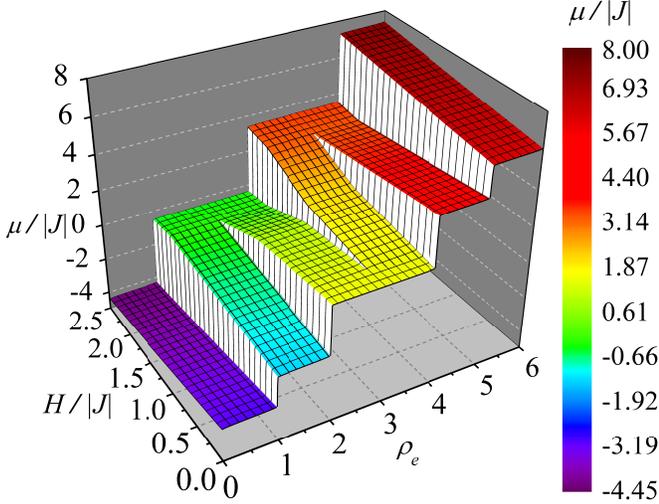}
\caption{\small (Color online) The chemical potential $\mu$ as a function of the electron density $\rho_e$ and the magnetic field $H$ for the Coulomb term $U/|J| = 3$, the kinetic parameter $t/|J| = 1.3$ and the temperature $k_{\rm B}T/|J|=0.015$.}
  \label{fig6}
\end{figure}

To complete the zero-temperature analysis, we report in this section the results obtained for the chemical potential. This quantity can be obtained either by means of the bisection method from Eq.~(\ref{eq:ro}), or by using the relation
\begin{eqnarray}
\label{eq:mu}
\mu = \left(\frac{\partial G}{\partial\rho_e}\right)_{\!T}.
\end{eqnarray}
The possibility of using two different ways to determine the chemical potential lies in the fact that $\mu$ and $\rho_e$ are conjugate variables. Typical behavior of the chemical potential as a function of the electron density $\rho_e$ and the magnetic field $H$ is presented in Fig.~\ref{fig6}. In order to contain all ground-state regions discussed in previous subsection, the Coulomb term and hopping parameter are fixed to the same values as in Fig.~\ref{fig5}(d). The temperature takes the lowest possible value for numerical calculations $k_{\rm B}T/|J|=0.015$. It is clear from Fig.~\ref{fig6} that the chemical potential corresponding to individual ground-state regions has a linear dependence on $H$, but it  remains constant during the $\rho_e$ change. The only exception is the region ${\rm R}_{2\to4}$, where the chemical potential does not change even when changing the field, nor changing the electron concentration in the model. Steep but continuous steps in 3D plot in Fig.~\ref{fig6} clearly reproduce real discontinuous jumps in $\mu(\rho_e)$ curves which exist between neighboring regions at the zero temperature.

\newpage

\subsection{Magnetization process}
\label{subsec:4C}

\begin{figure}[t!]
\vspace{0.25cm}
    \includegraphics[width = 1.0\columnwidth]{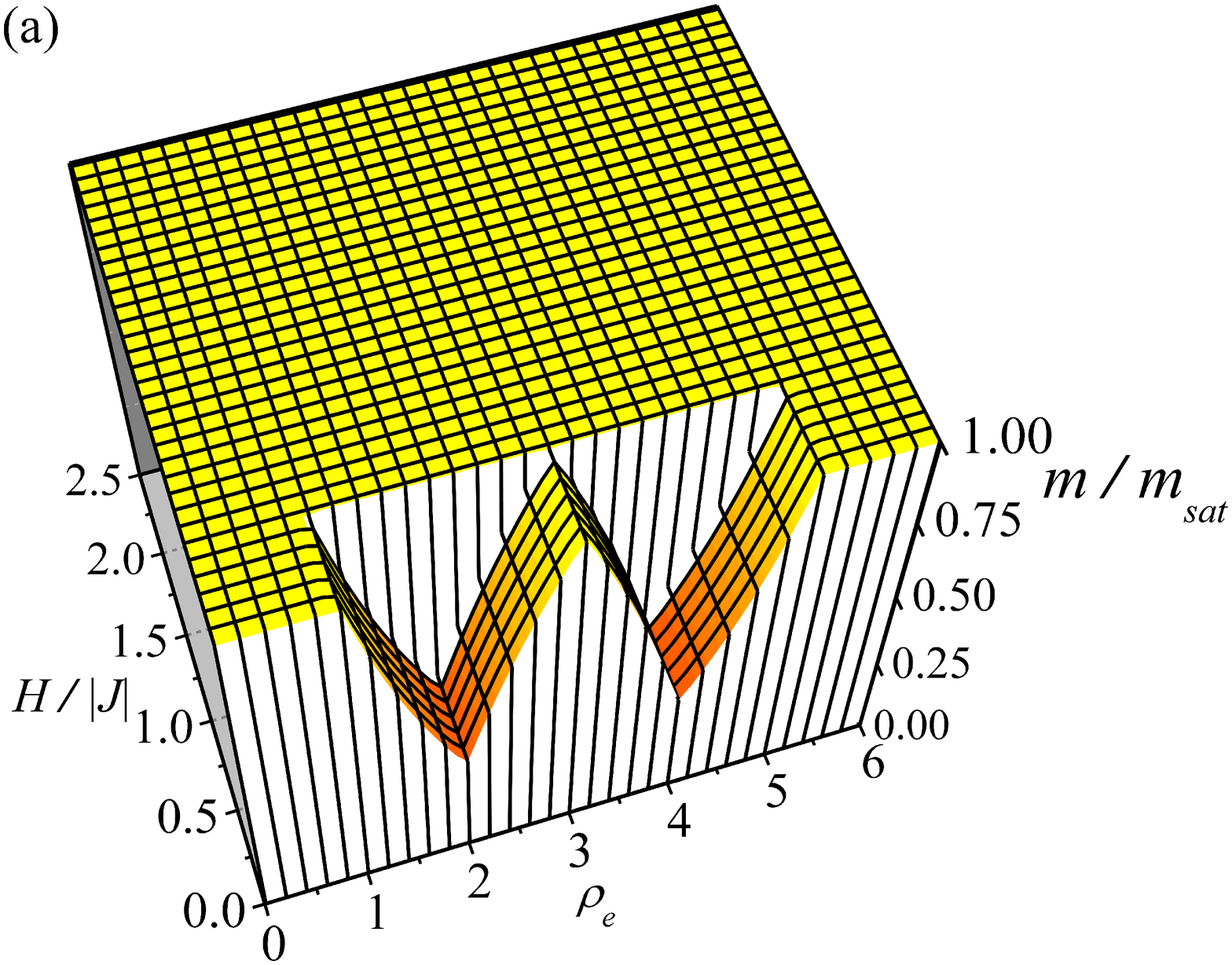}
     \\[0.4cm]
    \includegraphics[width = 1.0\columnwidth]{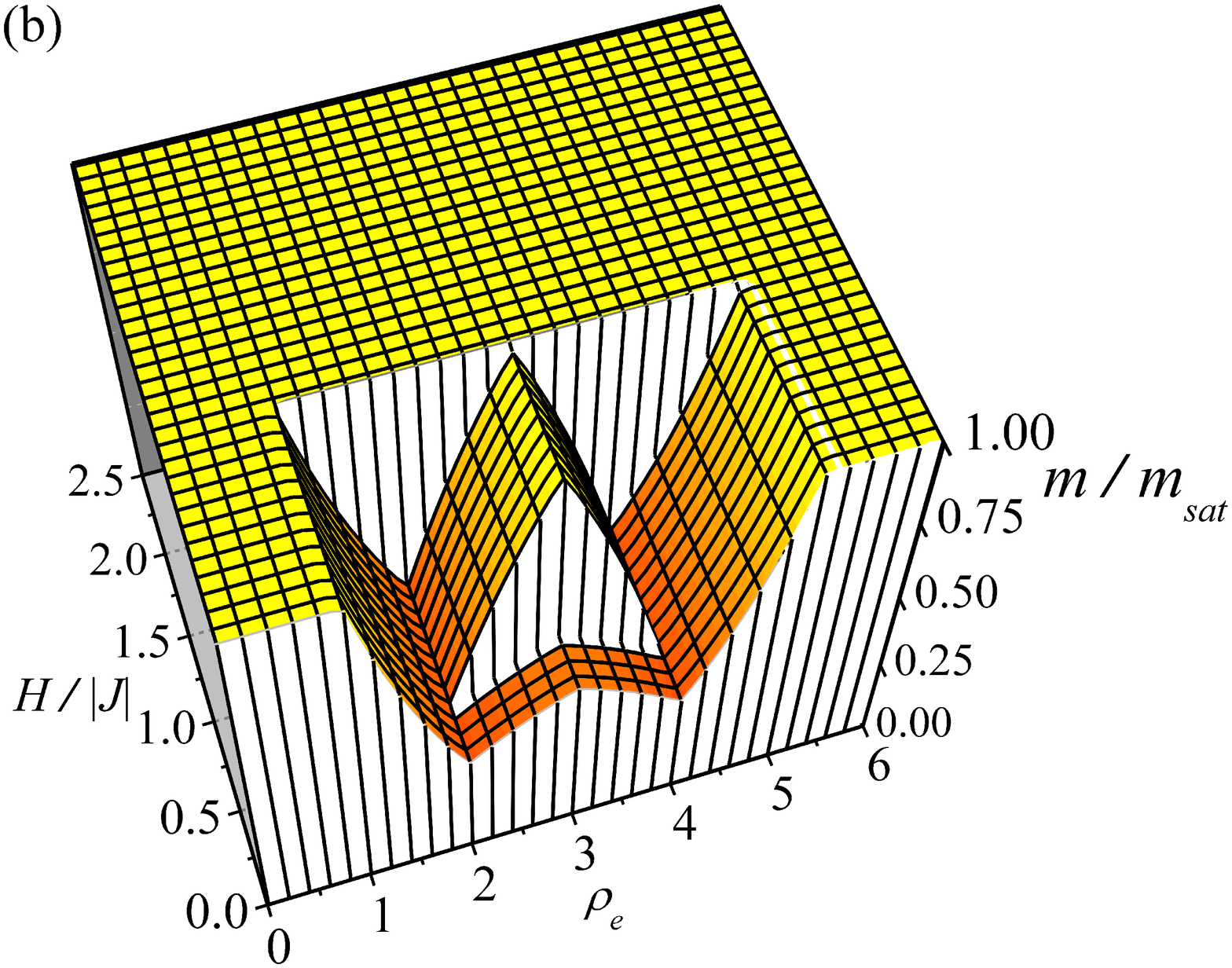}
    \\[0.4cm]
    \includegraphics[width = 1.0\columnwidth]{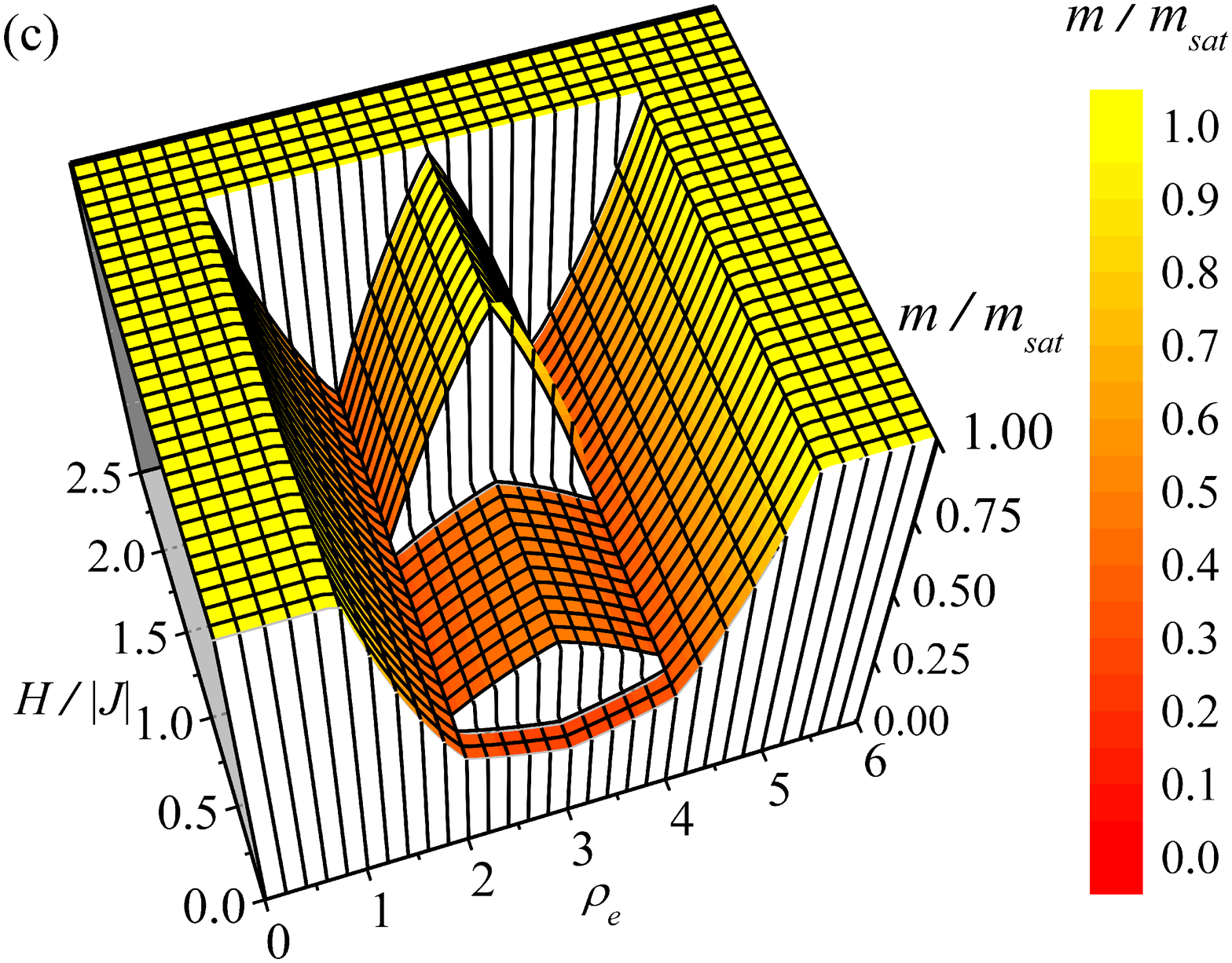}
\caption{\small (Color online) 3D views of the magnetization process for the Coulomb term $U/|J| = 3$ and the hopping parameters (a)~$t/|J| = 0.7$, (b)~$t/|J| = 0.9$, (c)~$t/|J| = 1.3$ at the  temperature $k_{\rm B}T/|J| = 0.015$.}
  \label{fig7}
\vspace{-0.5cm}
\end{figure}

As discussed in recent works~\cite{Cab01, Cab02, Rou06, Rou07, Str16}, a gradual change of the particle content in the model may lead to a formation of irrational plateaux in magnetization curves. In order to illustrate the existence of these plateaux in the
magnetization process of the studied spin-electron chain~(\ref{eq:H}), Fig.~\ref{fig7} shows three different 3D views of the low-temperature total magnetization normalized to its saturation value in the $\rho_e-H$ plane. It is noteworthy that results for the afore-mentioned magnetization has been realized by using the previously established relation~(\ref{eq:m}) in a combination with Eqs.~(\ref{eq:mImecn}). The reduced parameters $U/|J|$, $t/|J|$ take the same values as in Figs.~\ref{fig5}(b)--(d). Apart from a direct steep increase from the zero to saturation value in the zero-field limit, the displayed surfaces of the ratio $m/m_{sat}$ may include in total seven doping-dependent plateaux if $t>t_{b1}$. In particular, for relatively weak hopping terms $t_{b1}<t<t_{b2}$, four single plateaux can be observed in magnetization curves before reaching the saturation magnetization upon strengthening the applied magnetic field:
\begin{eqnarray}
\label{eq:mmsat1}
\frac{m}{m_{sat}} \!\!&=&\!\! \frac{3-\rho_e}{1+\rho_e} \qquad {\rm for}\,\,\, 1<\rho_e\leq2,\\
\label{eq:mmsat2}
\frac{m}{m_{sat}} \!\!&=&\!\! \frac{3\rho_e-5}{1+\rho_e} \quad\,\,\, {\rm for}\,\,\, 2<\rho_e\leq3,\\
\label{eq:mmsat3}
\frac{m}{m_{sat}} \!\!&=&\!\! \frac{13-3\rho_e}{7-\rho_e} \quad {\rm for}\,\,\, 3<\rho_e\leq4,\\
\frac{m}{m_{sat}} \!\!&=&\!\! \frac{\rho_e-3}{7-\rho_e} \qquad {\rm for}\,\,\, 4<\rho_e\leq5
\end{eqnarray}
[see Fig.~\ref{fig7}(b)].
Referring to Fig.~\ref{fig5}(b)--(d) one can conclude that the first plateau~(\ref{eq:mmsat1}) is pertinent to the region R$_{1\to2}$, second one~(\ref{eq:mmsat2}) corresponds to the region R$_{2\to3}$, while last two~(\ref{eq:mmsat3}) and~(\ref{eq:mmsat4}) reflect the existence of the regions R$_{3\to4}$ and R$_{4\to5}$, respectively. For stronger hopping terms $t_{b2}<t<t_{b3}$, the magnetization scenario includes other two plateaux that can be continuously tuned by electron doping according to the formulas
\begin{eqnarray}
\label{eq:mmsat3}
\frac{m}{m_{sat}} \!\!&=&\!\! \frac{\rho_e-1}{1+\rho_e} \qquad {\rm for}\,\,\, 2<\rho_e\leq3,
\\
\label{eq:mmsat4}
\frac{m}{m_{sat}} \!\!&=&\!\! \frac{5-\rho_e}{7-\rho_e} \qquad {\rm for}\,\,\, 3<\rho_e\leq4.
\end{eqnarray}
The plateau given by Eq.~(\ref{eq:mmsat3}) corresponds to the region R$_{2\to\widetilde{3}}$, while the plateau given by Eq.~(\ref{eq:mmsat4}) appears in magnetization curves when the region R$_{\widetilde{3}\to4}$ constitutes the ground state [compare Figs.~\ref{fig5}(c), (d) with Figs.~\ref{fig7}(c), (d)]. Last but not least, an intriguing plateau given by the condition
\begin{eqnarray}
\label{eq:m/msatR24}
\frac{m}{m_{sat}} =
\begin{cases}
\dfrac{1}{1+ \rho_e} &\quad {\rm for}\,\,\, 2 < \rho_e \leq 3,
\\[3mm]
\dfrac{1}{7 - \rho_e} &\quad {\rm for}\,\,\, 3 < \rho_e \leq 4\,
\end{cases}
\end{eqnarray}
can be observed in low-field parts of the magnetization curves if $t>t_{b3}$ [see Fig.~\ref{fig7}(d)]. Interestingly, this plateau is the only one, which has the non-monotonous dependence during doping process: for electron densities $2<\rho_e<3$, the value of the ratio $m/m_{sat}$ gradually decreases to its minimum $m/m_{sat}=1/4$ at $\rho_e=3$. On the other hand, the value of $m/m_{sat}$ increases with $\rho_e$ while for $3 < \rho_e \leq 4$. The observed plateau corresponds to the ground-state region~R$_{2\to4}$.

\section{Conclusions}
\label{sec:5}

The present work deals with a rigorously solvable double-tetrahedral chain, in which the Ising spins of the magnitude $\sigma=1/2$ localized at nodal lattice sites regularly alternate with triangular plaquettes occupied by mobile electrons. Ground-state properties and low-temperature magnetization process are examined in detail by assuming a variable electron density in the model.

It has been evidenced that the uniform change of the electron content in all triangular plaquettes controlled by the chemical potential $\mu$ in a combination with the competition between model parameters and the external magnetic field $H$ lead to the formation of one chiral and seven non-chiral phases in the zero-temperature $\mu-H$ parameter space. The macroscopic ground-state degeneracy arising from chiral degrees of freedom of mobile electrons is not suppressed by the external field in contrast to the one which appears due to a frustration of the localized Ising spins. Magnetization curves involve one or two fractional plateaux at one-third and/or one-half of the saturation magnetization, in addition to a direct steep increase from zero to saturation value found in the zero-field limit. On the other hand, a gradual electron doping results in eleven distinct regions at the absolute zero temperature. These regions can be distinguished from each other by an evolution of the possible electron distribution in triangular plaquettes during the doping process. It has been demonstrated that magnetization curves include in total seven irrational plateaux whose heights can be  continuously tuned by a gradual change of the electron content in the model.

It is well known that a steep variation of the magnetization at low magnetic fields points to a possible applicability for low-temperature magnetic refrigeration achieved through the enhanced magnetocaloric effect~\cite{Zhi03,Sch07,Sch13}. Considering this statement, the studied mixed-spin double-tetrahedral chain represents a quite good theoretical tool for a rigorous examination of the afore-mentioned phenomenon when the particle content is changed. Our research interest will continue in that direction.

\section*{ACKNOWLEDGMENT}
This work was financially supported by Ministry of Education, Science, Research and Sport of the Slovak Republic under the grant VEGA 1/0043/16 and by the Slovak Research and Development Agency under the contracts APVV-0097-12, APVV-16-0186.

\newpage

\begin{widetext}
\appendix
\appendixpageoff
\section{}
\label{app:1}
\begin{longtable*}{crlll}
\caption{Decomposition of electron configurations from $k$th triangular cluster into the orbits ${\cal O}_{|f_{kj}^{\rm ini}\rangle}$ of the cyclic symmetry group $C_{3}$ and the corresponding eigenenergies $E_{kl}$ of the block Hamiltonian~(\ref{eq:Hk}) for different numbers of mobile electrons $n_k=\{0,1,2,\ldots,6\}$. Here, we have introduced the following notations $h_{I}\! = H_{I}(\sigma_{k}^{z} + \sigma_{k+1}^{z})\!/2$, $h_{e} \!= J(\sigma_{k}^{z} + \sigma_{k+1}^{z}) - H_{e}$ and $\phi_{U,t} = \frac{1}{3} \arctan\left[9t\sqrt{U^4+27U^2t^2+243t^4}/U^{3}\right]$ in order to write the eigenenergies $E_{kl}$ in a more abbreviated form.}
\label{tab} \\
\noalign{\smallskip}\hline\hline\noalign{\smallskip}
$n_k$\,\, & $S_{k}^z$ \hspace{0.25cm} & $|f_{kj}^{\rm ini}\rangle$ & ${\cal O}_{|f_{kj}^{\rm ini}\rangle}$ & $E_{kl}$
\\
\noalign{\smallskip}\hline\hline\noalign{\smallskip}
\endfirsthead
\multicolumn{5}{l}%
{\tablename\ \thetable\ -- \textrm{Continued from previous page}} \\
\noalign{\smallskip}\hline\hline\noalign{\smallskip}
$n_k$ & $S_{k}^z$ \hspace{0.25cm} & $|f_{kj}^{\rm ini}\rangle$ & ${\cal O}_{|f_{kj}^{\rm ini}\rangle}$ & $E_{kl}$ \\
\noalign{\smallskip}\hline\hline\noalign{\smallskip}
\endhead
\hline \multicolumn{5}{r}{\textrm{Continued on next page}} \\
\endfoot
\noalign{\smallskip}\hline\hline\noalign{\smallskip}
\endlastfoot
$0$    & $0$  \hspace{0.25cm}  &  $|0\rangle$  &  $|0\rangle$    &  $-h_I$
\\
\noalign{\smallskip}\hline\noalign{\smallskip}
\multirow{1}{*}{$1$} & $-\frac{1}{2}$ \hspace{0.25cm} & $c_{k1,\downarrow}^{\dagger}|0\rangle$ &  $c_{k1,\downarrow}^{\dagger}|0\rangle$  & $-\mu-h_I-\frac{h_e}{2}-2t$
\\
&  &  &  $c_{k2,\downarrow}^{\dagger}|0\rangle$  & $-\mu-h_I-\frac{h_e}{2}+t$
\\
& &  &  $c_{k3,\downarrow}^{\dagger}|0\rangle$  & $-\mu-h_I-\frac{h_e}{2}+t$
\\
\noalign{\smallskip}\cline{2-5}\noalign{\smallskip}
\multirow{1}{*}{ } & $\frac{1}{2}$ \hspace{0.25cm} & $c_{k1,\uparrow}^{\dagger}|0\rangle$ &  ${c_{k1,\uparrow}^{\dagger}|0\rangle}$ & $-\mu-h_I+\frac{h_e}{2}-2t$
\\
&  &  &  $c_{k2,\uparrow}^{\dagger}|0\rangle$  & $-\mu-h_I+\frac{h_e}{2}+t$
\\
& &  &  $c_{k3,\uparrow}^{\dagger}|0\rangle$  & $-\mu-h_I+\frac{h_e}{2}+t$
\\
\noalign{\smallskip}\hline\noalign{\smallskip}
\multirow{1}{*}{$2$} & $-1$ \hspace{0.25cm} & $c_{k1,\downarrow}^{\dagger}c_{k2,\downarrow}^{\dagger}|0\rangle$  &  $c_{k1,\downarrow}^{\dagger}c_{k2,\downarrow}^{\dagger}|0\rangle$ &
$-2\mu-h_I-h_e+2t$
\\
&  &  &  $c_{k2,\downarrow}^{\dagger}c_{k3,\downarrow}^{\dagger}|0\rangle$  &
$-2\mu-h_I-h_e-t$
\\
&  &  &  $c_{k3,\downarrow}^{\dagger}c_{k1,\downarrow}^{\dagger}|0\rangle$  &
$-2\mu-h_I-h_e-t$
\\
\noalign{\smallskip}\cline{2-5}\noalign{\smallskip}
\multirow{1}{*}{ } & $0$ \hspace{0.25cm} & $c_{k1,\uparrow}^{\dagger}c_{k1,\downarrow}^{\dagger}|0\rangle$ &
	$c_{k1,\uparrow}^{\dagger}c_{k1,\downarrow}^{\dagger}|0\rangle$ &
	$-2\mu-h_I+2t$
	\\
&  &  &  $c_{k2,\uparrow}^{\dagger}c_{k3,\downarrow}^{\dagger}|0\rangle$ &
	$-2\mu-h_I-t+\frac{U}{2}\pm\frac{1}{2}\sqrt{(U+2t)^2+32t^2}$	
	\\
&  &  &  $c_{k2,\uparrow}^{\dagger}c_{k3,\downarrow}^{\dagger}|0\rangle$ &
\\
	\noalign{\smallskip}\cline{3-5}\noalign{\smallskip}
&     &
        $c_{k2,\uparrow}^{\dagger}c_{k3,\downarrow}^{\dagger}|0\rangle$ & $c_{k2,\uparrow}^{\dagger}c_{k3,\downarrow}^{\dagger}|0\rangle$
& $-2\mu-h_I-t$
\\
&   &      &   $c_{k3,\uparrow}^{\dagger}c_{k1,\downarrow}^{\dagger}|0\rangle$
& $-2\mu-h_I+\frac{t}{2}+\frac{U}{2}\pm\frac{1}{2}\sqrt{(U-t)^2+8t^2}$
\\
&   &      &   $c_{k1,\uparrow}^{\dagger}c_{k2,\downarrow}^{\dagger}|0\rangle$
&
\\
	\noalign{\smallskip}\cline{3-5}\noalign{\smallskip}
&     &
         $c_{k3,\downarrow}^{\dagger}c_{k1,\uparrow}^{\dagger}|0\rangle$ & $c_{k3,\downarrow}^{\dagger}c_{k1,\uparrow}^{\dagger}|0\rangle$
& $-2\mu-h_I-t$\\
&   &      &   $c_{k1,\downarrow}^{\dagger}c_{k2,\uparrow}^{\dagger}|0\rangle$
& $-2\mu-h_I+\frac{t}{2}+\frac{U}{2}\pm\frac{1}{2}\sqrt{(U-t)^2+8t^2}$
\\
&   &      &   $c_{k2,\downarrow}^{\dagger}c_{k3,\uparrow}^{\dagger}|0\rangle$
&
\\
  \noalign{\smallskip}\cline{2-5}\noalign{\smallskip}
\multirow{1}{*}{ } &  $1$ \hspace{0.25cm}  &
		$c_{k1,\uparrow}^{\dagger}c_{k2,\uparrow}^{\dagger}|0\rangle$  & $c_{k1,\uparrow}^{\dagger}c_{k2,\uparrow}^{\dagger}|0\rangle$  &
        $-2\mu-h_I+h_e+2t$
				\\
&   &  &  $c_{k2,\uparrow}^{\dagger}c_{k3,\uparrow}^{\dagger}|0\rangle$  &
        $-2\mu-h_I+h_e-t$ \\
&   &  &  $c_{k3,\uparrow}^{\dagger}c_{k1,\uparrow}^{\dagger}|0\rangle$  &
        $-2\mu-h_I+h_e-t$
				\\
\noalign{\smallskip}\hline\noalign{\smallskip}
\multirow{1}{*}{$3$} & $-\frac{3}{2}$ \hspace{0.25cm} &
        $c_{k1,\downarrow}^{\dagger}c_{k2,\downarrow}^{\dagger}c_{k3,\downarrow}^{\dagger}|0\rangle$ &
        $c_{k1,\downarrow}^{\dagger}c_{k2,\downarrow}^{\dagger}c_{k3,\downarrow}^{\dagger}|0\rangle$ & $-3\mu-h_I-\frac{3h_e}{2}$
				\\
   \noalign{\smallskip}\cline{2-5}\noalign{\smallskip}
\multirow{1}{*}{} & $-\frac{1}{2}$ \hspace{0.25cm} &
        $c_{k1,\downarrow}^{\dagger}c_{k2,\downarrow}^{\dagger}c_{k3,\uparrow}^{\dagger}|0\rangle$ &
        $c_{k1,\downarrow}^{\dagger}c_{k2,\downarrow}^{\dagger}c_{k3,\uparrow}^{\dagger}|0\rangle$ & $-3\mu-h_I-\frac{h_e}{2}$\\
&   &
                                                                                                   &
        $c_{k2,\downarrow}^{\dagger}c_{k3,\downarrow}^{\dagger}c_{k1,\uparrow}^{\dagger}|0\rangle$ & $-3\mu-h_I-\frac{h_e}{2}+U$\\
&   &
                                                                                                   &
        $c_{k3,\downarrow}^{\dagger}c_{k1,\downarrow}^{\dagger}c_{k2,\uparrow}^{\dagger}|0\rangle$ & $-3\mu-h_I-\frac{h_e}{2}+U$\\	
    \noalign{\smallskip}\cline{3-5}\noalign{\smallskip}
&   &
        $c_{k2,\uparrow}^{\dagger}c_{k2,\downarrow}^{\dagger}c_{k3,\downarrow}^{\dagger}|0\rangle$ &
        $c_{k2,\uparrow}^{\dagger}c_{k2,\downarrow}^{\dagger}c_{k3,\downarrow}^{\dagger}|0\rangle$ & $-3\mu-h_I-\frac{h_e}{2}+\frac{2U}{3}-\frac{2}{3}\sqrt{U^2+27t^2}\cos\phi$\\
&   &
                                                                                                   &
        $c_{k3,\uparrow}^{\dagger}c_{k3,\downarrow}^{\dagger}c_{k1,\downarrow}^{\dagger}|0\rangle$ & $-3\mu-h_I-\frac{h_e}{2}+\frac{2U}{3}\,+$\\
&   &
                                                                                                   &
        $c_{k1,\uparrow}^{\dagger}c_{k1,\downarrow}^{\dagger}c_{k2,\downarrow}^{\dagger}|0\rangle$ &$\hspace{6mm}+\frac{1}{3}\sqrt{U^2+27t^2}\left(\cos\phi\pm\sqrt{3}\sin\phi\right)$\\

     \noalign{\smallskip}\cline{3-5}\noalign{\smallskip}
&	&
        $c_{k3,\downarrow}^{\dagger}c_{k1,\uparrow}^{\dagger}c_{k1,\downarrow}^{\dagger}|0\rangle$ &
        $c_{k3,\downarrow}^{\dagger}c_{k1,\uparrow}^{\dagger}c_{k1,\downarrow}^{\dagger}|0\rangle$ & $-3\mu-h_I-\frac{h_e}{2}+\frac{2U}{3}-\frac{2}{3}\sqrt{U^2+27t^2}\cos\phi$\\
&   &
                                                                                                   &
        $c_{k1,\downarrow}^{\dagger}c_{k2,\uparrow}^{\dagger}c_{k2,\downarrow}^{\dagger}|0\rangle$ & $-3\mu-h_I-\frac{h_e}{2}+\frac{2U}{3}\,+$\\
&   &
                                                                                                   &
        $c_{k2,\downarrow}^{\dagger}c_{k3,\uparrow}^{\dagger}c_{k3,\downarrow}^{\dagger}|0\rangle$ &$\hspace{6mm}+\frac{1}{3}\sqrt{U^2+27t^2}\left(\cos\phi\pm\sqrt{3}\sin\phi\right)$\\
   \noalign{\smallskip}\cline{2-5}\noalign{\smallskip}
 \multirow{1}{*}{} & $\frac{1}{2}$ \hspace{0.25cm} &
         $c_{k1,\uparrow}^{\dagger}c_{k2,\uparrow}^{\dagger}c_{k3,\downarrow}^{\dagger}|0\rangle$ &
        $c_{k1,\uparrow}^{\dagger}c_{k2,\uparrow}^{\dagger}c_{k3,\downarrow}^{\dagger}|0\rangle$ & $-3\mu-h_I+\frac{h_e}{2}$\\
&   &
                                                                                                   &
        $c_{k2,\uparrow}^{\dagger}c_{k3,\uparrow}^{\dagger}c_{k1,\downarrow}^{\dagger}|0\rangle$ & $-3\mu-h_I+\frac{h_e}{2}+U$\\
&   &
                                                                                                   &
        $c_{k3,\uparrow}^{\dagger}c_{k1,\uparrow}^{\dagger}c_{k2,\downarrow}^{\dagger}|0\rangle$ & $-3\mu-h_I+\frac{h_e}{2}+U$\\
	\noalign{\smallskip}\cline{3-5}\noalign{\smallskip}
&	&
        $c_{k2,\uparrow}^{\dagger}c_{k2,\downarrow}^{\dagger}c_{k3,\uparrow}^{\dagger}|0\rangle$ &
        $c_{k2,\uparrow}^{\dagger}c_{k2,\downarrow}^{\dagger}c_{k3,\uparrow}^{\dagger}|0\rangle$ & $-3\mu-h_I+\frac{h_e}{2}+\frac{2U}{3}-\frac{2}{3}\sqrt{U^2+27t^2}\cos\phi$\\
&   &
                                                                                                   &
        $c_{k3,\uparrow}^{\dagger}c_{k3,\downarrow}^{\dagger}c_{k1,\uparrow}^{\dagger}|0\rangle$ & $-3\mu-h_I+\frac{h_e}{2}+\frac{2U}{3}\,+$\\
&   &
                                                                                                   &
        $c_{k1,\uparrow}^{\dagger}c_{k1,\downarrow}^{\dagger}c_{k2,\uparrow}^{\dagger}|0\rangle$ &$\hspace{6mm}+\frac{1}{3}\sqrt{U^2+27t^2}\left(\cos\phi\pm\sqrt{3}\sin\phi\right)$\\
	\noalign{\smallskip}\cline{3-5}\noalign{\smallskip}
&   &
        $c_{k3,\uparrow}^{\dagger}c_{k1,\uparrow}^{\dagger}c_{k1,\downarrow}^{\dagger}|0\rangle$ &
        $c_{k3,\uparrow}^{\dagger}c_{k1,\uparrow}^{\dagger}c_{k1,\downarrow}^{\dagger}|0\rangle$ & $-3\mu-h_I+\frac{h_e}{2}+\frac{2U}{3}-\frac{2}{3}\sqrt{U^2+27t^2}\cos\phi$\\
&   &
                                                                                                   &
        $c_{k1,\uparrow}^{\dagger}c_{k2,\uparrow}^{\dagger}c_{k2,\downarrow}^{\dagger}|0\rangle$ & $-3\mu-h_I+\frac{h_e}{2}+\frac{2U}{3}\,+$\\
&   &
                                                                                                   &
        $c_{k2,\uparrow}^{\dagger}c_{k3,\uparrow}^{\dagger}c_{k3,\downarrow}^{\dagger}|0\rangle$ &$\hspace{6mm}+\frac{1}{3}\sqrt{U^2+27t^2}\left(\cos\phi\pm\sqrt{3}\sin\phi\right)$\\
    \noalign{\smallskip}\cline{2-5}\noalign{\smallskip}
\multirow{1}{*}{ } & $\frac{3}{2}$ \hspace{0.25cm} &
        $c_{k1,\uparrow}^{\dagger}c_{k2,\uparrow}^{\dagger}c_{k3,\uparrow}^{\dagger}|0\rangle$ &
        $c_{k1,\uparrow}^{\dagger}c_{k2,\uparrow}^{\dagger}c_{k3,\uparrow}^{\dagger}|0\rangle$ & $-3\mu-h_I+\frac{3h_e}{2}$\\
        \noalign{\smallskip}\hline\noalign{\smallskip}
\multirow{1}{*}{$4$} & $-1$ \hspace{0.25cm} & $c_{k1,\uparrow}^{\dagger}c_{k1,\downarrow}^{\dagger}c_{k2,\downarrow}^{\dagger}c_{k3,\downarrow}^{\dagger}|0\rangle$  &  $c_{k1,\uparrow}^{\dagger}c_{k1,\downarrow}^{\dagger}c_{k2,\downarrow}^{\dagger}c_{k3,\downarrow}^{\dagger}|0\rangle$ &
$-4\mu-h_I-h_e+2t+U$
\\
& &  &    $c_{k2,\uparrow}^{\dagger}c_{k2,\downarrow}^{\dagger}c_{k3,\downarrow}^{\dagger}c_{k1,\downarrow}^{\dagger}|0\rangle$  &
$-4\mu-h_I-h_e-t+U$
\\
& &  &    $c_{k3,\uparrow}^{\dagger}c_{k3,\downarrow}^{\dagger}c_{k1,\downarrow}^{\dagger}c_{k2,\downarrow}^{\dagger}|0\rangle$  &
$-4\mu-h_I-h_e-t+U$
\\
\noalign{\smallskip}\cline{2-5}\noalign{\smallskip}
\multirow{1}{*}{ } & $0$ \hspace{0.25cm} & $c_{k1,\uparrow}^{\dagger}c_{k1,\downarrow}^{\dagger}c_{k2,\uparrow}^{\dagger}c_{k2,\downarrow}^{\dagger}|0\rangle$ &
	 $c_{k1,\uparrow}^{\dagger}c_{k1,\downarrow}^{\dagger}c_{k2,\uparrow}^{\dagger}c_{k2,\downarrow}^{\dagger}|0\rangle$ &
	$-4\mu-h_I+2t+U$
	\\
& &   &   $c_{k2,\uparrow}^{\dagger}c_{k2,\downarrow}^{\dagger}c_{k3,\uparrow}^{\dagger}c_{k3,\downarrow}^{\dagger}|0\rangle$ &
	$-4\mu-h_I-t+\frac{3U}{2}\pm\frac{1}{2}\sqrt{(U+2t)^2+32t^2}$	
	\\
& &  &  $c_{k3,\uparrow}^{\dagger}c_{k3,\downarrow}^{\dagger}c_{k1,\uparrow}^{\dagger}c_{k1,\downarrow}^{\dagger}|0\rangle$
\\
	\noalign{\smallskip}\cline{3-5}\noalign{\smallskip}
&   &
        $c_{k2,\uparrow}^{\dagger}c_{k2,\downarrow}^{\dagger}c_{k3,\uparrow}^{\dagger}c_{k1,\downarrow}^{\dagger}|0\rangle$ & $c_{k2,\uparrow}^{\dagger}c_{k2,\downarrow}^{\dagger}c_{k3,\uparrow}^{\dagger}c_{k1,\downarrow}^{\dagger}|0\rangle$
& $-4\mu-h_I-t+U$
\\
& &      &    $c_{k3,\uparrow}^{\dagger}c_{k3,\downarrow}^{\dagger}c_{k1,\uparrow}^{\dagger}c_{k2,\downarrow}^{\dagger}|0\rangle$
& $-4\mu-h_I+\frac{t}{2}+\frac{3U}{2}\pm\frac{1}{2}\sqrt{(U-t)^2+8t^2}$
\\
&      &   & $c_{k1,\uparrow}^{\dagger}c_{k1,\downarrow}^{\dagger}c_{k2,\uparrow}^{\dagger}c_{k3,\downarrow}^{\dagger}|0\rangle$
\\
	\noalign{\smallskip}\cline{3-5}\noalign{\smallskip}
 &   &
         $c_{k3,\uparrow}^{\dagger}c_{k3,\downarrow}^{\dagger}c_{k1,\downarrow}^{\dagger}c_{k2,\uparrow}^{\dagger}|0\rangle$ & $c_{k3,\uparrow}^{\dagger}c_{k3,\downarrow}^{\dagger}c_{k1,\downarrow}^{\dagger}c_{k2,\uparrow}^{\dagger}|0\rangle$
& $-4\mu-h_I-t+U$\\
& &      &   $c_{k1,\uparrow}^{\dagger}c_{k1,\downarrow}^{\dagger}c_{k2,\downarrow}^{\dagger}c_{k3,\uparrow}^{\dagger}|0\rangle$
& $-4\mu-h_I+\frac{t}{2}+\frac{3U}{2}\pm\frac{1}{2}\sqrt{(U-t)^2+8t^2}$
\\
& &      &   $c_{k2,\uparrow}^{\dagger}c_{k2,\downarrow}^{\dagger}c_{k3,\downarrow}^{\dagger}c_{k1,\uparrow}^{\dagger}|0\rangle$
&
\\
   \noalign{\smallskip}\cline{2-5}\noalign{\smallskip}
\multirow{1}{*}{ } & $1$ \hspace{0.25cm} &
		 $c_{k1,\uparrow}^{\dagger}c_{k1,\downarrow}^{\dagger}c_{k2,\uparrow}^{\dagger}c_{k3,\uparrow}^{\dagger}|0\rangle$  & $c_{k1,\uparrow}^{\dagger}c_{k1,\downarrow}^{\dagger}c_{k2,\uparrow}^{\dagger}c_{k3,\uparrow}^{\dagger}|0\rangle$  &
        $-4\mu-h_I+h_e+2t+U$
				\\
& &  &  $c_{k2,\uparrow}^{\dagger}c_{k2,\downarrow}^{\dagger}c_{k3,\uparrow}^{\dagger}c_{k1,\uparrow}^{\dagger}|0\rangle$  &
        $-4\mu-h_I+h_e-t+U$ \\
& &  &  $c_{k3,\uparrow}^{\dagger}c_{k3,\downarrow}^{\dagger}c_{k1,\uparrow}^{\dagger}c_{k2,\uparrow}^{\dagger}|0\rangle$  &
        $-4\mu-h_I+h_e-t+U$
				\\
\noalign{\smallskip}\hline\noalign{\smallskip}
\multirow{1}{*}{$5$} & $-\frac{1}{2}$ \hspace{0.25cm} & $c_{k1,\downarrow}^{\dagger}c_{k2,\uparrow}^{\dagger}c_{k2,\downarrow}^{\dagger}c_{k3,\uparrow}^{\dagger}c_{k3,\downarrow}^{\dagger}|0\rangle$ &  $c_{k1,\downarrow}^{\dagger}c_{k2,\uparrow}^{\dagger}c_{k2,\downarrow}^{\dagger}c_{k3,\uparrow}^{\dagger}c_{k3,\downarrow}^{\dagger}|0\rangle$  & $-5\mu-h_I-\frac{h_e}{2}-2t+2U$
\\
&  &  &  $c_{k2,\downarrow}^{\dagger}c_{k3,\uparrow}^{\dagger}c_{k3,\downarrow}^{\dagger}c_{k1,\uparrow}^{\dagger}c_{k1,\downarrow}^{\dagger}|0\rangle$  & $-5\mu-h_I-\frac{h_e}{2}+t+2U$
\\
& &  &  $c_{k3,\downarrow}^{\dagger}c_{k1,\uparrow}^{\dagger}c_{k1,\downarrow}^{\dagger}c_{k2,\uparrow}^{\dagger}c_{k2,\downarrow}^{\dagger}|0\rangle$  & $-5\mu-h_I-\frac{h_e}{2}+t+2U$
\\
\noalign{\smallskip}\cline{2-5}\noalign{\smallskip}
\multirow{1}{*}{ } & $\frac{1}{2}$ \hspace{0.25cm} & $c_{k1,\uparrow}^{\dagger}c_{k2,\uparrow}^{\dagger}c_{k2,\downarrow}^{\dagger}c_{k3,\uparrow}^{\dagger}c_{k3,\downarrow}^{\dagger}|0\rangle$ &  $c_{k1,\uparrow}^{\dagger}c_{k2,\uparrow}^{\dagger}c_{k2,\downarrow}^{\dagger}c_{k3,\uparrow}^{\dagger}c_{k3,\downarrow}^{\dagger}|0\rangle$ & $-5\mu-h_I+\frac{h_e}{2}-2t+2U$
\\
&  &  &  $c_{k2,\uparrow}^{\dagger}c_{k3,\uparrow}^{\dagger}c_{k3,\downarrow}^{\dagger}c_{k1,\uparrow}^{\dagger}c_{k1,\downarrow}^{\dagger}|0\rangle$  & $-5\mu-h_I+\frac{h_e}{2}+t+2U$
\\
&  &  &  $c_{k3,\uparrow}^{\dagger}c_{k1,\uparrow}^{\dagger}c_{k1,\downarrow}^{\dagger}c_{k2,\uparrow}^{\dagger}c_{k2,\downarrow}^{\dagger}|0\rangle$  & $-5\mu-h_I+\frac{h_e}{2}+t+2U$
\\
\noalign{\smallskip}\hline\noalign{\smallskip}
\multirow{1}{*}{6} & $0$ \hspace{0.25cm} & $c_{k1,\uparrow}^{\dagger}c_{k1,\downarrow}^{\dagger}c_{k2,\uparrow}^{\dagger}c_{k2,\downarrow}^{\dagger}c_{k3,\uparrow}^{\dagger}c_{k3,\downarrow}^{\dagger}|0\rangle$ &  $c_{k1,\uparrow}^{\dagger}c_{k1,\downarrow}^{\dagger}c_{k2,\uparrow}^{\dagger}c_{k2,\downarrow}^{\dagger}c_{k3,\uparrow}^{\dagger}c_{k3,\downarrow}^{\dagger}|0\rangle$ & $-6\mu-h_I+3U$
\end{longtable*}

\section{}
\label{app:2}
The list of state vectors indicating eigenstates of mobile electrons in the $k$th triangular plaquette that create the eigenvectors of the ground-state phases~(\ref{eq:S_0,2,4,6})--(\ref{eq:S_3chiral}) and~(\ref{eq:S_2,4chiral}) :
\begin{eqnarray}
|n_k =0, S_k^{z} = 0\rangle \!\!&=&\!\! |0\rangle,
\\
\Big|n_k =1, S_k^{z} = {\textstyle-\dfrac{1}{2}}\Big\rangle \!\!&=&\!\! \frac{1}{\sqrt{3}}\left(
c_{k1,\downarrow}^{\dagger} + c_{k2,\downarrow}^{\dagger} + c_{k3,\downarrow}^{\dagger}\right)\!|0\rangle,
\\
\Big|n_k =1, S_k^{z} ={\textstyle\dfrac{1}{2}}\Big\rangle \!\!&=&\!\! \frac{1}{\sqrt{3}}\left(
c_{k1,\uparrow}^{\dagger} + c_{k2,\uparrow}^{\dagger} + c_{k3,\uparrow}^{\dagger}\right)\!|0\rangle,
\\
\label{eq:n2ch-}
|n_k =2, S_k^{z} = -1\rangle_{L,R} \!\!&=&\!\!
\begin{cases}
\dfrac{1}{\sqrt{3}}\left(
c_{k1,\downarrow}^{\dagger}c_{k2,\downarrow}^{\dagger} + {\rm e}^{\frac{2\pi\,{\rm i}}{3}}c_{k2,\downarrow}^{\dagger}c_{k3,\downarrow}^{\dagger} + {\rm e}^{\frac{4\pi\,{\rm i}}{3}}c_{k3,\downarrow}^{\dagger}c_{k1,\downarrow}^{\dagger}\right)\!|0\rangle,
           {}
\\[3mm]
\dfrac{1}{\sqrt{3}}\left(
c_{k1,\downarrow}^{\dagger}c_{k2,\downarrow}^{\dagger} + {\rm e}^{\frac{4\pi\,{\rm i}}{3}}c_{k2,\downarrow}^{\dagger}c_{k3,\downarrow}^{\dagger} + {\rm e}^{\frac{2\pi\,{\rm i}}{3}}c_{k3,\downarrow}^{\dagger}c_{k1,\downarrow}^{\dagger}\right)\!|0\rangle,  {}
\end{cases}
\end{eqnarray}
\begin{eqnarray}
\label{eq:n2}
|n_k =2, S_k^{z} = 0\rangle \!\!&=&\!\!
{\cal A}_{2} \!\left(c_{k1,\uparrow}^{\dag}c_{k2,\downarrow}^{\dag}\!+ c_{k2,\uparrow}^{\dag}c_{k3,\downarrow}^{\dag} \!+ c_{k3,\uparrow}^{\dag}c_{k1,\downarrow}^{\dag} -\, c_{k1,\downarrow}^{\dag}c_{k2,\uparrow}^{\dag}\! - c_{k2,\downarrow}^{\dag}c_{k3,\uparrow}^{\dag} \!
 - c_{k3,\downarrow}^{\dag}c_{k1,\uparrow}^{\dag}\right)\!|0\rangle
\nonumber\\ & &
 +\,  {\cal B}_{2} \sum_{j=1}^{3}c_{kj,\uparrow}^{\dag}c_{kj,\downarrow}^{\dag}|0\rangle,{}
\\
\label{eq:n2ch+}
|n_k =2, S_k^{z} = 1\rangle_{L,R} \!\!&=&\!\!
\begin{cases}
\dfrac{1}{\sqrt{3}}\left(
c_{k1,\uparrow}^{\dagger}c_{k2,\uparrow}^{\dagger} + {\rm e}^{\frac{2\pi\,{\rm i}}{3}}c_{k2,\uparrow}^{\dagger}c_{k3,\uparrow}^{\dagger} + {\rm e}^{\frac{4\pi\,{\rm i}}{3}}c_{k3,\uparrow}^{\dagger}c_{k1,\uparrow}^{\dagger}\right)\!|0\rangle,
           {}
\\[3mm]
\dfrac{1}{\sqrt{3}}\left(
c_{k1,\downarrow}^{\dagger}c_{k2,\downarrow}^{\dagger} + {\rm e}^{\frac{4\pi\,{\rm i}}{3}}c_{k2,\downarrow}^{\dagger}c_{k3,\downarrow}^{\dagger} + {\rm e}^{\frac{2\pi\,{\rm i}}{3}}c_{k3,\downarrow}^{\dagger}c_{k1,\downarrow}^{\dagger}\right)\!|0\rangle,           {},
\end{cases}
\\
\Big|n_k =3, S_k^{z} = {\textstyle-\dfrac{3}{2}}\Big\rangle \!\!&=&\!\!
c_{k1,\downarrow}^{\dagger}c_{k2,\downarrow}^{\dagger}c_{k3,\downarrow}^{\dagger}|0\rangle,
\\[2mm]
\label{AppB:S_3a}
\Big|n_k =3, S_k^{z} = {\textstyle-\dfrac{1}{2}}\Big\rangle_{L,R} \!\!&=&\!\!
\begin{cases}
\,{\cal A}_{3}\! \left(c_{k1,\uparrow}^{\dagger}c_{k1,\downarrow}^{\dagger}c_{k2,\downarrow}^{\dagger}
             +{\rm e}^{\frac{2\pi\,{\rm i}}{3}}c_{k2,\uparrow}^{\dagger}c_{k2,\downarrow}^{\dagger}c_{k3,\downarrow}^{\dagger}
            +{\rm e}^{\frac{4\pi\,{\rm i}}{3}}c_{k3,\uparrow}^{\dagger}c_{k3,\downarrow}^{\dagger}c_{k1,\downarrow}^{\dagger}\right)\!|0\rangle{}
\\[2mm]
    \hspace{4mm}+\,
    {\cal B}_{3}^{+}c_{k1,\downarrow}^{\dagger}c_{k2,\downarrow}^{\dagger}c_{k3,\uparrow}^{\dagger}|0\rangle
   +{\cal B}_{3}^{-}c_{k2,\downarrow}^{\dagger}c_{k3,\downarrow}^{\dagger}c_{k1,\uparrow}^{\dagger}|0\rangle -\left({\cal B}_{3}^{+}+{\cal B}_{3}^{-}\right) c_{k3,\downarrow}^{\dagger}c_{k1,\downarrow}^{\dagger}c_{k2,\uparrow}^{\dagger}|0\rangle
  \\[2mm]
    \hspace{4mm}+\,
  {\cal C}_{3}^{-}c_{k1,\downarrow}^{\dagger}c_{k2,\uparrow}^{\dagger}c_{k2,\downarrow}^{\dagger}|0\rangle
   +{\cal C}_{3}^{+}c_{k2,\downarrow}^{\dagger}c_{k3,\uparrow}^{\dagger}c_{k3,\downarrow}^{\dagger}|0\rangle -\left({\cal C}_{3}^{+}+{\cal C}_{3}^{-}\right) c_{k3,\downarrow}^{\dagger}c_{k1,\uparrow}^{\dagger}c_{k1,\downarrow}^{\dagger}|0\rangle
           {}
\\[2mm]
\,{\cal A}_{3}\! \left(c_{k1,\uparrow}^{\dagger}c_{k1,\downarrow}^{\dagger}c_{k2,\downarrow}^{\dagger}
             +{\rm e}^{\frac{4\pi\,{\rm i}}{3}}c_{k2,\uparrow}^{\dagger}c_{k2,\downarrow}^{\dagger}c_{k3,\downarrow}^{\dagger}
            +{\rm e}^{\frac{2\pi\,{\rm i}}{3}}c_{k3,\uparrow}^{\dagger}c_{k3,\downarrow}^{\dagger}c_{k1,\downarrow}^{\dagger}\right)\!|0\rangle{}
\\[2mm]
    \hspace{4mm}+\,
    {\cal B}_{3}^{+}c_{k1,\downarrow}^{\dagger}c_{k2,\downarrow}^{\dagger}c_{k3,\uparrow}^{\dagger}|0\rangle
   +{\cal B}_{3}^{-}c_{k2,\downarrow}^{\dagger}c_{k3,\downarrow}^{\dagger}c_{k1,\uparrow}^{\dagger}|0\rangle -\left({\cal B}_{3}^{+}+{\cal B}_{3}^{-}\right) c_{k3,\downarrow}^{\dagger}c_{k1,\downarrow}^{\dagger}c_{k2,\uparrow}^{\dagger}|0\rangle
  \\[2mm]
    \hspace{4mm}+\,
  {\cal C}_{3}^{-}c_{k1,\downarrow}^{\dagger}c_{k2,\uparrow}^{\dagger}c_{k2,\downarrow}^{\dagger}|0\rangle
   +{\cal C}_{3}^{+}c_{k2,\downarrow}^{\dagger}c_{k3,\uparrow}^{\dagger}c_{k3,\downarrow}^{\dagger}|0\rangle -\left({\cal C}_{3}^{+}+{\cal C}_{3}^{-}\right) c_{k3,\downarrow}^{\dagger}c_{k1,\uparrow}^{\dagger}c_{k1,\downarrow}^{\dagger}|0\rangle  {},
\end{cases}
\\
\label{AppB:S_3b}
\Big|n_k =3, S_k^{z} = {\textstyle\dfrac{1}{2}}\Big\rangle_{L,R} \!\!&=&\!\!
\begin{cases}
\,
{\cal A}_{3}\! \left(c_{k1,\uparrow}^{\dagger}c_{k1,\downarrow}^{\dagger}c_{k2,\uparrow}^{\dagger}
             +{\rm e}^{\frac{2\pi\,{\rm i}}{3}}c_{k2,\uparrow}^{\dagger}c_{k2,\downarrow}^{\dagger}c_{k3,\uparrow}^{\dagger}
            +{\rm e}^{\frac{4\pi\,{\rm i}}{3}}c_{k3,\uparrow}^{\dagger}c_{k3,\downarrow}^{\dagger}c_{k1,\uparrow}^{\dagger}\right)\!|0\rangle{}
\\[2mm]
    \hspace{4mm}+\,
    {\cal B}_{3}^{+}c_{k1,\uparrow}^{\dagger}c_{k2,\uparrow}^{\dagger}c_{k3,\downarrow}^{\dagger}|0\rangle
   +{\cal B}_{3}^{-}c_{k2,\uparrow}^{\dagger}c_{k3,\uparrow}^{\dagger}c_{k1,\downarrow}^{\dagger}|0\rangle -\left({\cal B}_{3}^{+}+{\cal B}_{3}^{-}\right) c_{k3,\uparrow}^{\dagger}c_{k1,\uparrow}^{\dagger}c_{k2,\downarrow}^{\dagger}|0\rangle
  \\[2mm]
    \hspace{4mm}+\,
  {\cal C}_{3}^{-}c_{k1,\uparrow}^{\dagger}c_{k2,\uparrow}^{\dagger}c_{k2,\downarrow}^{\dagger}|0\rangle
   +{\cal C}_{3}^{+}c_{k2,\uparrow}^{\dagger}c_{k3,\uparrow}^{\dagger}c_{k3,\downarrow}^{\dagger}|0\rangle -\left({\cal C}_{3}^{+}+{\cal C}_{3}^{-}\right) c_{k3,\uparrow}^{\dagger}c_{k1,\uparrow}^{\dagger}c_{k1,\downarrow}^{\dagger}|0\rangle
{}
\\[2mm]
\,
{\cal A}_{3}\! \left(c_{k1,\uparrow}^{\dagger}c_{k1,\downarrow}^{\dagger}c_{k2,\uparrow}^{\dagger}
             +{\rm e}^{\frac{4\pi\,{\rm i}}{3}}c_{k2,\uparrow}^{\dagger}c_{k2,\downarrow}^{\dagger}c_{k3,\uparrow}^{\dagger}
            +{\rm e}^{\frac{2\pi\,{\rm i}}{3}}c_{k3,\uparrow}^{\dagger}c_{k3,\downarrow}^{\dagger}c_{k1,\uparrow}^{\dagger}\right)\!|0\rangle{}
\\[2mm]
    \hspace{4mm}+\,
    {\cal B}_{3}^{+}c_{k1,\uparrow}^{\dagger}c_{k2,\uparrow}^{\dagger}c_{k3,\downarrow}^{\dagger}|0\rangle
   +{\cal B}_{3}^{-}c_{k2,\uparrow}^{\dagger}c_{k3,\uparrow}^{\dagger}c_{k1,\downarrow}^{\dagger}|0\rangle -\left({\cal B}_{3}^{+}+{\cal B}_{3}^{-}\right) c_{k3,\uparrow}^{\dagger}c_{k1,\uparrow}^{\dagger}c_{k2,\downarrow}^{\dagger}|0\rangle
  \\[2mm]
    \hspace{4mm}+\,
  {\cal C}_{3}^{-}c_{k1,\uparrow}^{\dagger}c_{k2,\uparrow}^{\dagger}c_{k2,\downarrow}^{\dagger}|0\rangle
   +{\cal C}_{3}^{+}c_{k2,\uparrow}^{\dagger}c_{k3,\uparrow}^{\dagger}c_{k3,\downarrow}^{\dagger}|0\rangle -\left({\cal C}_{3}^{+}+{\cal C}_{3}^{-}\right) c_{k3,\uparrow}^{\dagger}c_{k1,\uparrow}^{\dagger}c_{k1,\downarrow}^{\dagger}|0\rangle
{},
\end{cases}
\\
\Big|n_k =3, S_k^{z} = {\textstyle\dfrac{3}{2}}\Big\rangle \!\!&=&\!\! c_{k1,\uparrow}^{\dagger}c_{k2,\uparrow}^{\dagger}c_{k3,\uparrow}^{\dagger}|0\rangle,
\\[1mm]
\label{eq:n4ch-}
|n_k =4, S_k^{z} = -1\rangle_{L,R} \!\!&=&\!\!
\begin{cases}
\dfrac{1}{\sqrt{3}}\left(
c_{k1,\downarrow}^{\dagger}c_{k2,\downarrow}^{\dagger}c_{k3,\uparrow}^{\dagger}c_{k3,\downarrow}^{\dagger} + {\rm e}^{\frac{2\pi\,{\rm i}}{3}}c_{k2,\downarrow}^{\dagger}c_{k3,\downarrow}^{\dagger}c_{k1,\uparrow}^{\dagger}c_{k1,\downarrow}^{\dagger} + {\rm e}^{\frac{4\pi\,{\rm i}}{3}}c_{k3,\downarrow}^{\dagger}c_{k1,\downarrow}^{\dagger}c_{k2,\uparrow}^{\dagger}c_{k2,\downarrow}^{\dagger}\right)\!|0\rangle,
           {}
\\[3mm]
\dfrac{1}{\sqrt{3}}\left(
c_{k1,\downarrow}^{\dagger}c_{k2,\downarrow}^{\dagger}c_{k3,\uparrow}^{\dagger}c_{k3,\downarrow}^{\dagger} + {\rm e}^{\frac{4\pi\,{\rm i}}{3}}c_{k2,\downarrow}^{\dagger}c_{k3,\downarrow}^{\dagger}c_{k1,\uparrow}^{\dagger}c_{k1,\downarrow}^{\dagger} + {\rm e}^{\frac{2\pi\,{\rm i}}{3}}c_{k3,\downarrow}^{\dagger}c_{k1,\downarrow}^{\dagger}c_{k2,\uparrow}^{\dagger}c_{k2,\downarrow}^{\dagger}\right)\!|0\rangle,        {},
\end{cases}
\\
\label{eq:n4}
|n_k =4, S_k^{z} = 0\rangle \!\!&=&\!\! {\cal A}_{4} \Big(c_{k1,\uparrow}^{\dag}c_{k1,\downarrow}^{\dag}c_{k2,\uparrow}^{\dag}c_{k3,\downarrow}^{\dag}\!+ c_{k2,\uparrow}^{\dag}c_{k2,\downarrow}^{\dag}c_{k3,\uparrow}^{\dag}c_{k1,\downarrow}^{\dag}\!+ c_{k3,\uparrow}^{\dag}c_{k3,\downarrow}^{\dag}c_{k1,\uparrow}^{\dag}c_{k2,\downarrow}^{\dag} -\, c_{k1,\uparrow}^{\dag}c_{k1,\downarrow}^{\dag}c_{k2,\downarrow}^{\dag}c_{k3,\uparrow}^{\dag}
\nonumber \\
& & \,\, -\, c_{k2,\uparrow}^{\dag}c_{k2,\downarrow}^{\dag}c_{k3,\downarrow}^{\dag}c_{k1,\uparrow}^{\dag} \!
-
c_{k3,\uparrow}^{\dag}c_{k3,\downarrow}^{\dag}c_{k1,\downarrow}^{\dag}c_{k2,\uparrow}^{\dag}\Big)|0\rangle +  {\cal B}_{4} \sum_{j=1}^{3}c_{kj,\uparrow}^{\dag}c_{kj,\downarrow}^{\dag}c_{k(j+1),\uparrow}^{\dag}c_{k(j+1),\downarrow}^{\dag}|0\rangle,
\hspace{10mm}
\\
\label{eq:n4ch+}
|n_k =4, S_k^{z} = 1\rangle_{L,R} \!\!&=&\!\!
\begin{cases}
\dfrac{1}{\sqrt{3}}\left(
c_{k1,\uparrow}^{\dagger}c_{k2,\uparrow}^{\dagger}c_{k3,\uparrow}^{\dagger}c_{k3,\downarrow}^{\dagger} + {\rm e}^{\frac{2\pi\,{\rm i}}{3}}c_{k2,\uparrow}^{\dagger}c_{k3,\uparrow}^{\dagger}c_{k1,\uparrow}^{\dagger}c_{k1,\downarrow}^{\dagger} + {\rm e}^{\frac{4\pi\,{\rm i}}{3}}c_{k3,\uparrow}^{\dagger}c_{k1,\uparrow}^{\dagger}c_{k2,\uparrow}^{\dagger}c_{k2,\downarrow}^{\dagger}\right)\!|0\rangle,
           {}
\\[3mm]
\dfrac{1}{\sqrt{3}}\left(
c_{k1,\uparrow}^{\dagger}c_{k2,\uparrow}^{\dagger}c_{k3,\uparrow}^{\dagger}c_{k3,\downarrow}^{\dagger} + {\rm e}^{\frac{4\pi\,{\rm i}}{3}}c_{k2,\uparrow}^{\dagger}c_{k3,\uparrow}^{\dagger}c_{k1,\uparrow}^{\dagger}c_{k1,\downarrow}^{\dagger} + {\rm e}^{\frac{2\pi\,{\rm i}}{3}}c_{k3,\uparrow}^{\dagger}c_{k1,\uparrow}^{\dagger}c_{k2,\uparrow}^{\dagger}c_{k2,\downarrow}^{\dagger}\right)\!|0\rangle,       {}
\end{cases}
\\
\Big|n_k =5, S_k^{z} = {\textstyle-\dfrac{1}{2}}\Big\rangle \!\!&=&\!\! \frac{1}{\sqrt{3}}\left(
c_{k1,\downarrow}^{\dagger}c_{k2,\uparrow}^{\dagger}c_{k2,\downarrow}^{\dagger}c_{k3,\uparrow}^{\dagger}c_{k3,\downarrow}^{\dagger} + c_{k2,\downarrow}^{\dagger}c_{k3,\uparrow}^{\dagger}c_{k3,\downarrow}^{\dagger}c_{k1,\uparrow}^{\dagger}c_{k1,\downarrow}^{\dagger} + c_{k3,\downarrow}^{\dagger}c_{k1,\uparrow}^{\dagger}c_{k1,\downarrow}^{\dagger}c_{k2,\uparrow}^{\dagger}c_{k2,\downarrow}^{\dagger}\right)\!|0\rangle,
\\
\Big|n_k =5, S_k^{z} = {\textstyle\dfrac{1}{2}}\Big\rangle \!\!&=&\!\! \frac{1}{\sqrt{3}}\left(
c_{k1,\uparrow}^{\dagger}c_{k2,\uparrow}^{\dagger}c_{k2,\downarrow}^{\dagger}c_{k3,\uparrow}^{\dagger}c_{k3,\downarrow}^{\dagger} + c_{k2,\uparrow}^{\dagger}c_{k3,\uparrow}^{\dagger}c_{k3,\downarrow}^{\dagger}c_{k1,\uparrow}^{\dagger}c_{k1,\downarrow}^{\dagger} + c_{k3,\uparrow}^{\dagger}c_{k1,\uparrow}^{\dagger}c_{k1,\downarrow}^{\dagger}c_{k2,\uparrow}^{\dagger}c_{k2,\downarrow}^{\dagger}\right)\!|0\rangle,
\\
|n_k =6, S_k^{z} = 0\rangle \!\!&=&\!\!  c_{k1,\uparrow}^{\dagger}c_{k1,\downarrow}^{\dagger}c_{k2,\uparrow}^{\dagger}c_{k2,\downarrow}^{\dagger}c_{k3,\uparrow}^{\dagger}c_{k3,\downarrow}^{\dagger}|0\rangle,
\end{eqnarray}
where ${\rm i}=\sqrt{-1}$ and the coefficients ${\cal A}_{2-4}$, ${\cal B}_{2}$, ${\cal B}_{3}^{\pm}$, ${\cal B}_{4}$  and ${\cal C}_{3}^{\pm}$ are:
\begin{eqnarray*}
{\cal A}_{2} \!\!&=&\!\!  \frac{1}{\sqrt{6}}\frac{4\mu+U+2t+\sqrt{(U+2t)^{2}+32t^{2}}}{\sqrt{\big(4\mu+U+2t+\sqrt{(U+2t)^{2}+32t^{2}}\,\big)^{\!2} + 32t^2}}\,,
\\
{\cal A}_{3} \!\!&=&\!\! \frac{1}{\sqrt{6}}\frac{\left[9\mu - 2U + 2\sqrt{U^2+27t^2}\cos\left(\phi_{U,t}\right)\right]\!\left[9\mu + U + 2\sqrt{U^2+27t^2}\cos\left(\phi_{U,t}\right)\right] - 27t^2}{\sqrt{\left[9\mu - 2U + 2\sqrt{U^2+27t^2}\cos\left(\phi_{U,t}\right)\right]^2\!\left[9\mu + U + 2\sqrt{U^2+27t^2}\cos\left(\phi_{U,t}\right)\right]^2 + 243t^2U^2 + 2187t^4}}\,,
\\
{\cal A}_{4} \!\!&=&\!\!  \frac{1}{\sqrt{6}}\frac{8\mu+U+2t+\sqrt{(U+2t)^{2}+32t^{2}}}{\sqrt{\big(8\mu+U+2t+\sqrt{(U+2t)^{2}+32t^{2}}\,\big)^{\!2} + 32t^2}}\,,
\\
{\cal B}_{2} \!\!&=&\!\!  \frac{1}{\sqrt{6}}\frac{8t}{\sqrt{\big(4\mu+U+2t+\sqrt{(U+2t)^{2}+32t^{2}}\,\big)^{\!2} + 32t^2}}\,,
\\
{\cal B}_{3}^{\pm} \!\!&=&\!\! \pm\frac{9t}{\sqrt{6}}\frac{9\mu + U + 2\sqrt{U^2+27t^2}\cos\left(\phi_{U,t}\right) \pm 3t}{\sqrt{\left[9\mu - 2U + 2\sqrt{U^2+27t^2}\cos\left(\phi_{U,t}\right)\right]^2\!\left[9\mu + U + 2\sqrt{U^2+27t^2}\cos\left(\phi_{U,t}\right)\right]^2 + 243t^2U^2 + 2187t^4}}\,,
\\
{\cal B}_{4} \!\!&=&\!\! -\frac{1}{\sqrt{6}}\frac{8t}{\sqrt{\big(8\mu+U+2t+\sqrt{(U+2t)^{2}+32t^{2}}\,\big)^{\!2} + 32t^2}}\,,
\\
{\cal C}_{3}^{\pm} \!\!&=&\!\! \pm\frac{9t}{\sqrt{6}}\frac{9\mu - 2U + 2\sqrt{U^2+27t^2}\cos\left(\phi_{U,t}\right) \pm 3t}{\sqrt{\left[9\mu - 2U + 2\sqrt{U^2+27t^2}\cos\left(\phi_{U,t}\right)\right]^2\!\left[9\mu + U + 2\sqrt{U^2+27t^2}\cos\left(\phi_{U,t}\right)\right]^2 + 243t^2U^2 + 2187t^4}}\,.
\end{eqnarray*}

\section{}
\label{app:3}
The list of analytical expressions for the first-order phase transitions between individual ground-state phases of the model~(\ref{eq:H}), which have been obtained by comparing the ground-state energies~(\ref{eq:S_0,2,4,6})--(\ref{eq:S_3chiral}):
\begin{eqnarray}
\begin{array}{lll}
{\rm S}_0\!-\!{\rm S}_1&:&H = J - 2\mu - 4t,
\\[2mm]
{\rm S}_1\!-\!{\rm S}_2&:&H = J + 2\mu - U - 2t +\!\sqrt{(U + 2t)^2\!+32t^2}
\qquad ({\rm for}\,\, t>t_{b1}),
\nonumber \\[2mm]
{\rm S}_1\!-\!{\rm S}_3&:&H = J - 2\mu + 2t,
\nonumber \\[2mm]
{\rm S}_2\!-\!{\rm S}_3&:&H = J - \dfrac{2\mu}{3} -\dfrac{U}{3} + \dfrac{2t}{3} +\dfrac{1}{3}\sqrt{(U + 2t)^2\!+32t^2}
\qquad ({\rm for}\,\,t>t_{b1}),
\nonumber \\[2mm]
{\rm S}_2\!-\!{\rm \widetilde{S}}_3&:&H = J - 2\mu +\dfrac{U}{3} + 2t + \!\sqrt{(U + 2t)^2\!+32t^2} -\dfrac{4}{3}\sqrt{U^{2} + 27t^{2}}\,\cos\left(\phi_{U,\,t}\right)
\qquad ({\rm for}\,\,t>t_{b2}),
\nonumber \\[2mm]
{\rm S}_2\!-\!{\rm S}_4&:&\mu = \dfrac{U}{2}
\qquad (t>t_{b3}),
\nonumber \\[2mm]
{\rm \widetilde{S}}_3\!-\!{\rm S}_3&:&H = J - \dfrac{2U}{3} + \dfrac{2}{3}\sqrt{U^{2} + 27t^{2}}\,\cos\left(\phi_{U,\,t}\right) \qquad ({\rm for}\,\,t>t_{b2}),
\nonumber \\[2mm]
{\rm \widetilde{S}}_3\!-\!{\rm S}_4&:&H = J + 2\mu -\dfrac{5U}{3} + 2t + \!\sqrt{(U + 2t)^2\!+32t^2} -\dfrac{4}{3}\sqrt{U^{2} + 27t^{2}}\,\cos\left(\phi_{U,\,t}\right)
\qquad ({\rm for}\,\,t>t_{b2}),
\nonumber \\[2mm]
{\rm S}_3\!-\!{\rm S}_4&:&H = J + \dfrac{2\mu}{3} -U + \dfrac{2t}{3} +\dfrac{1}{3}\sqrt{(U + 2t)^2\!+32t^2}
\qquad ({\rm for}\,\,t>t_{b1}),
\nonumber \\[2mm]
{\rm S}_3\!-\!{\rm S}_5&:&H = J + 2\mu - 2U + 2t,
\nonumber \\[2mm]
{\rm S}_4\!-\!{\rm S}_5&:&H = J - 2\mu + U - 2t +\sqrt{(U + 2t)^2\!+32t^2}
\qquad ({\rm for}\,\,t>t_{b1}),
\nonumber \\[2mm]
{\rm S}_5\!-\!{\rm S}_6&:&H = J + 2\mu - 2U - 4t.
\nonumber
\end{array}
\end{eqnarray}
In above, $t_{b1}$,  $t_{b2}$ and $t_{b3}$ are given by Eqs.~(\ref{eq:tb1}), (\ref{eq:tb2}) and (\ref{eq:tb3}), respectively.
\end{widetext}

\end{document}